\documentclass[12pt,preprint]{aastex}
\slugcomment{} 
\shorttitle{}
\shortauthors{}
\begin{document}
%%%%%%%%%%%%%%%%%%%%%%%%%%%%%%%%%%%%%%%%%%%%%%%%%%%%%%%%%%%%%%%%%%%%%%
%%%%%%%%%%%%%%%%%%%%%%%%%%%%%%%%%%%%%%%%%%%%%%%%%%%%%%%%%%%%%%%%%%%%%%
\title{Subaru Weak Lensing Measurements of Four Strong Lensing
  Clusters: Are Lensing Clusters Over-Concentrated?\altaffilmark{1}}    
%%%%%%%%%%%%%%%%%%%%%%%%%%%%%%%%%%%%%%%%%%%%%%%%%%%%%%%%%%%%%%%%%%%%%%
%%%%%%%%%%%%%%%%%%%%%%%%%%%%%%%%%%%%%%%%%%%%%%%%%%%%%%%%%%%%%%%%%%%%%%
%
%%%%%%%%%%%%%%%%%%%%%%%%%%%%%%%%%%%%%%%%%%%%%%%%%%%%%%%%%%%%%%%%%%%%%%
\author{
Masamune Oguri,\altaffilmark{2} 
Joseph F. Hennawi,\altaffilmark{3} 
Michael D. Gladders,\altaffilmark{4} 
H{\aa}kon Dahle,\altaffilmark{5} \\
Priyamvada Natarajan,\altaffilmark{6} 
Neal Dalal,\altaffilmark{7} 
Benjamin P. Koester,\altaffilmark{4,8} \\
Keren Sharon,\altaffilmark{4,8,9} and
Matthew Bayliss\altaffilmark{4,8}
}

\altaffiltext{1}{Based on data collected at
Subaru Telescope, which is operated by the National Astronomical
  Observatory of Japan. Based on observations obtained at the Gemini
  Observatory, which is operated by the Association of Universities
  for Research in Astronomy, Inc., under a cooperative agreement with
  the NSF on behalf of the Gemini partnership: the National Science
  Foundation (United States), the Science and Technology Facilities
  Council (United Kingdom), the National Research Council (Canada),
  CONICYT (Chile), the Australian Research Council (Australia),
  Minist\'{e}rio da Ci\^{e}ncia e Tecnologia (Brazil) and SECYT 
(Argentina)} 
\altaffiltext{2}{Kavli Institute for Particle Astrophysics and 
Cosmology, Stanford University, 2575 Sand Hill Road, Menlo Park, 
CA 94025, USA.}
\altaffiltext{3}{Department of Astronomy, University of California 
Berkeley, Berkeley, CA 94720, USA.}
\altaffiltext{4}{Department of Astronomy \& Astrophysics, 
University of Chicago, Chicago, IL 60637, USA.}
\altaffiltext{5}{Institute of Theoretical Astrophysics, University of 
Oslo, P.O. Box 1029, Blindern, N-0315 Oslo, Norway.}
\altaffiltext{6}{Department of Astronomy, Yale University, P. O. Box
  208101, New Haven CT 06511-208101, USA.}
\altaffiltext{7}{Canadian Institute for Theoretical Astrophysics, 
60 St. George Street, University of Toronto, Toronto, ON M5S3H8, Canada.}
\altaffiltext{8}{Kavli Institute for Cosmological Physics, 
The University of Chicago, Chicago, IL 60637, USA.}
\altaffiltext{9}{School of Physics and Astronomy, Tel-Aviv 
University, Tel-Aviv 69978, Israel.}

\begin{abstract}
We derive radial mass profiles of four strong lensing selected
clusters which show prominent giant arcs (Abell 1703, SDSS J1446+3032,
SDSS J1531+3414, and SDSS J2111$-$0115), by combining detailed strong
lens modeling with weak lensing shear measured from deep {\it Subaru}
Suprime-cam images. Weak lensing signals are detected at high
significance for all four clusters, whose redshifts range from
$z=0.28$ to $0.64$. We demonstrate that adding strong lensing
information with known arc redshifts significantly improves
constraints on the mass density profile, compared to those obtained
from weak lensing alone. While the mass profiles are well fitted by
the universal form predicted in $N$-body simulations of the
$\Lambda$-dominated cold dark matter model, all four clusters appear
to be slightly more centrally concentrated (the concentration
parameters $c_{\rm vir}\sim 8$) than theoretical predictions, even
after accounting for the bias toward higher concentrations inherent in
lensing selected samples. Our results are consistent with previous
studies which similarly detected a concentration excess, and increases
the total number of clusters studied with the combined strong and weak
lensing technique to ten. Combining our sample with previous work, we
find that clusters with larger Einstein radii are more anomalously
concentrated. We also present a detailed model of the lensing cluster
Abell 1703 with constraints from multiple image families, and find
the dark matter inner density profile to be cuspy with the slope
consistent with $-1$, in agreement with expectations.    
\end{abstract}


%%%%%%%%%%%%%%%%%%%%%%%%%%%%%%%%%%%%%%%%%%%%%%%%
%%%%%%%%%%%%%%%%%%%%%%%%%%%%%%%%%%%%%%%%%%%%%%%%
%%%%%%%%%%%%%%%%%%%%%%%%%%%%%%%%%%%%%%%%%%%%%%%%
\section{Introduction}
%%%%%%%%%%%%%%%%%%%%%%%%%%%%%%%%%%%%%%%%%%%%%%%%
%%%%%%%%%%%%%%%%%%%%%%%%%%%%%%%%%%%%%%%%%%%%%%%%
%%%%%%%%%%%%%%%%%%%%%%%%%%%%%%%%%%%%%%%%%%%%%%%%

The current standard model of structure formation is successful in
explaining various cosmological observations, such as the cosmic
microwave background \citep{spergel03}, the clustering pattern of
galaxies \citep{eisenstein05}, the Ly$\alpha$ forest fluctuations in
the intergalactic medium \citep{mcdonald05}, and the abundance of
clusters of galaxies \citep{gladders07,vikhlinin09}. In the theory,
the growth of structure is driven by the gravitational instability of
dark matter. The initial density fluctuations are nearly
scale-invariant and Gaussian, as inferred from the cosmic microwave
background observations \citep[e,g.,][]{komatsu09}. The spectrum of
the fluctuation, together with the ``cold'' ansatz of dark matter,
suggests that astronomical objects form through the bottom-up assembly
process.    

Clusters of galaxies serve as one of the most important tests of the
standard structure formation model. Clusters are the largest
virialized objects in the universe. Their high virial temperatures
suggest that most of baryon remain hot in massive clusters and
dissipative cooling of baryons is less efficient. Therefore, the
density profile of clusters can well be approximated by the
distribution of dark matter which is sensitive to both the nature of
dark matter and the dark matter assemble history. In the standard
collisionless cold dark matter (CDM) scenario, the density profile of
dark matter found in $N$-body simulations has a universal form with
its slope progressively shallower toward the center \citep[][hereafter
  NFW]{navarro97}. Cluster-scale halos are predicted to have rather
low-concentration mass profiles \citep[e.g.,][]{bullock01,neto07}. 
These are important theoretical predictions that should be confronted  
with observations.   

Gravitational lensing is suited for this purpose, because it probes
the distribution of matter directly, regardless of the light
distribution. Weak gravitational lensing, which takes advantage of
small distortions of background galaxies to reconstruct the mass
distribution, is particularly powerful in extracting outer mass
profiles of massive clusters. On the other hand, strong lensing
provides robust measurements of cluster masses near the
center. Therefore, it is essential to combine strong and weak lensing
analysis in order to constrain density profiles over a wide range of 
radii \citep{natarajan98,natarajan02,bradac06,bradac08a,bradac08b,
diego07,limousin07,hicks07,deb08,merten09}. 

Cluster mass profiles grow in importance, particularly given possible
high concentrations  found in some lens-rich clusters.
\citet{broadhurst05a} argued that the radial mass profile of Abell
1689 (A1689) from combined strong and weak lensing analysis appears to
be significantly more centrally concentrated than expected from
$N$-body simulations of dark matter 
\citep[see also][]{clowe01,king02,halkola06,bardeau07,limousin07,umetsu08}.
Such high concentrations have also been suggested in other lensing
clusters, such as CL0024 \citep{kneib03} and MS2137 \citep{gavazzi03}.
A possible resolution is to consider a triaxial cluster with its 
major axis aligned with the line-of-sight direction
\citep{clowe04,oguri05,gavazzi05,corless07}. Because of the lensing
bias that lens-rich clusters are more likely to be oriented with the
major axis \citep{hennawi07}, the high concentration of A1689 can in
practice be marginally reconciled with the $\Lambda$CDM predictions
\citep{oguri05,hennawi07,oguri09,corless09}. The broad range of apparent
concentrations expected in the $\Lambda$CDM model, both from the
intrinsic scatter of concentrations and from additional scatter due to
the projection effect, indicates the need for the statistical study of
concentrations using a well-controlled sample of lensing clusters
\citep[e.g.,][]{broadhurst08a}.   

In this paper, we present new detailed lensing analysis of four
clusters. All the clusters are new strong lensing clusters discovered
by our survey of giant arcs \citep{hennawi08} using a catalog of
massive clusters constructed from the imaging and spectroscopic data
of the Sloan Digital Sky Survey \citep[SDSS;][]{york00}. We exploit
wide field-of-view and excellent image quality of the {\it Subaru}
Suprime-cam \citep{miyazaki02} to determine the outer mass profiles of
the clusters with weak lensing technique. The weak lensing signals are
combined with strong lensing constraints with redshifts of arcs
obtained from the {\it Gemini} telescope. The resulting radial mass
profiles of these clusters constrained from strong and weak lensing
are compared with the $\Lambda$CDM prediction.

The structure of this paper is as follows. We summarize our cluster
sample for analysis and follow-up data in \S\ref{sec:data}. Then we
perform strong and weak lensing analyses in \S\ref{sec:slens} and
\S\ref{sec:wlens}, respectively. These results are combined in
\S\ref{sec:s+w}. Our results are summarized in \S\ref{sec:summary}.
In Appendix~\ref{sec:gnfw}, we discuss the inner density profile of
one of our target clusters, Abell 1703, constrained from strong lens
modeling. Throughout the paper, we adopt a cosmological model with the
matter density $\Omega_M=0.26$, the cosmological constant
$\Omega_\Lambda=0.74$, and the dimensionless Hubble constant $h=0.72$. 

%%%%%%%%%%%%%%%%%%%%%%%%%%%%%%%%%%%%%%%%%%%%%%%%
%%%%%%%%%%%%%%%%%%%%%%%%%%%%%%%%%%%%%%%%%%%%%%%%
%%%%%%%%%%%%%%%%%%%%%%%%%%%%%%%%%%%%%%%%%%%%%%%%
\section{Cluster Sample and Follow-up Data}
\label{sec:data}
%%%%%%%%%%%%%%%%%%%%%%%%%%%%%%%%%%%%%%%%%%%%%%%%
%%%%%%%%%%%%%%%%%%%%%%%%%%%%%%%%%%%%%%%%%%%%%%%%
%%%%%%%%%%%%%%%%%%%%%%%%%%%%%%%%%%%%%%%%%%%%%%%%

We selected our targets for detailed lensing analysis from our sample
of lensing clusters with definitive arcs in the SDSS Giant Arc Survey
\citep[SGAS;][]{hennawi08}. The cluster lens sample has been
constructed as follows. We take advantage of a sample of the richest
clusters from the $8000\,{\rm deg^2}$ photometric data of the SDSS
using a red-sequence cluster finding algorithm
\citep{gladders00,gladders05}. We then obtained follow-up  images of
the richest clusters with the Wisconsin Indiana Yale NOAO (WIYN) 3.5m
telescope, the University of Hawaii 88-inch telescope (UH88), and the
Nordic Optical Telescope (NOT), and checked the images visually to
locate new giant arcs. Thus the SGAS probes the comoving volume of 
$\sim 6$~Gpc$^3$, representing the largest giant arc survey conducted
to date. Although the survey is still ongoing, it has already
uncovered more than 30 new lensing clusters that exhibit giant
arcs. See \citet{hennawi08} for more details on the survey method and
initial results of the SGAS.  

In this paper, we study the following four clusters; Abell 1703
(hereafter A1703, $z=0.281$), SDSS J1446+3032 (SDSS1446, $z=0.464$),
SDSS J1531+3414 (SDSS1531, $z=0.335$), and SDSS J2111$-$0115
(SDSS2111, $z=0.637$). Giant arcs of all the four clusters were newly
discovered by our survey \citep{hennawi08}. \citet{limousin08},
\citet{saha09}, and \citet{richard09} conducted strong lensing
analysis of A1703 based on the {\it Hubble Space Telescope}
images. \citet{broadhurst08a} presented results of mass modeling of
A1703 from both strong and weak lensing. Besides these, no lensing
analysis has been published for these clusters.   

%%%%%%%%%%%%%%%%%%%%%%%%%%%%%%%%%%%%%%%%%%%%%%%%
\subsection{Imaging Follow-up}
%%%%%%%%%%%%%%%%%%%%%%%%%%%%%%%%%%%%%%%%%%%%%%%%

We conducted imaging observations of these clusters with the
Suprime-Cam \citep{miyazaki02} on the Subaru 8.2-meter telescope on
2007 June 15. It covers a field of view $\sim 34'\times27'$ with the
pixel scale of $0\farcs202$. For each cluster, we obtained deep $g$-,
$r$-, and $i$-band images. We assign longer exposure time to $r$-band
images because we conduct weak lensing analysis using $r$-band
images. The $g$- and $i$-band images are taken to study colors of
strong lens candidates and to select galaxies for shear measurements. 
We summarize the exposure time, the seeing size, and the limiting
magnitude of each image in Table~\ref{tab:images}. The imaging data
are reduced using SDFRED \citep{yagi02,ouchi04}. The photometric
calibration is performed using the SDSS data by comparing magnitudes
of stars between SDSS and Subaru images. Astrometric calibration is
performed using reference objects in the USNO-A2.0 catalog
\citep{monet98}. The resulting astrometric accuracy is
$\sim0\farcs4$. We construct object catalogs for these images  using
SExtractor \citep{bertin96}. In this paper we basically use total
magnitudes (MAG\_AUTO). However, for measurements of galaxies colors
(\S\ref{sec:wlsrc}) we adopt aperture magnitudes (MAG\_APER) with an
aperture diameter of $2''$, because aperture magnitudes tend to
provide better measurements of colors. All the magnitudes are
corrected for Galactic extinction \citep{schlegel98}.   

%%%%%%%%%%%%%%%%%%%%%%%%%%%%%%%%%%%%%%%%%%%%%%%%
\subsection{Spectroscopic Follow-up}
\label{sec:arcspec}
%%%%%%%%%%%%%%%%%%%%%%%%%%%%%%%%%%%%%%%%%%%%%%%%

Spectroscopic observations were conducted for three of the four
clusters studied in this paper. SDSS1446, SDSS1531, and SDSS2111 were
observed with the Frederick C. Gillett Telescope (Gemini North)
between the months of 2008 February and 2008 July.  The primary goal
of the spectroscopic observations was to obtain redshifts of arcs to
facilitate strong lensing modeling. We briefly summarize the
spectroscopy here, and refer the reader to Hennawi et al. (2009, in
preparation) for additional details.   

All spectroscopic observations were carried out using the Gemini
Multi-Object Spectrograph \citep[GMOS;][]{hook04} using multi-object
slitmasks in microslit nod-and-shuffle (N\&S) mode 
\citep[e.g.,][]{glazebrook01,abraham04}.  Custom slitmasks were
designed and arcs were targeted based on their colors in the deep
Subaru imaging.  After targeting all of the arc candidates, any
remaining slits were placed on cluster members, easily identified by
their red sequence colors.

Spectra were taken with the R150\_G5306 grating in first order which
gives a dispersion of $3.5$~{\AA} per pixel, with six pixels per
resolution element resulting in a spectral FWHM$\simeq 940\,{\rm
  km\,s^{-1}}$. Although the R150 grating offers broad spectral range
  from the atmospheric cutoff to $\lambda \gtrsim 1\mu{\rm m}$, 
the drop in sensitivity at the blue and red extremes, due both to the
GMOS CCD and the R150 grating efficiency, results in effective
spectral coverage of $4000-9500$\AA.  Our exposure times were 2400~sec
resulting in 1200~sec effective integration for each of the two
submasks. Three exposures were taken for each target. Thus if an arc
was targeted on both submasks (typical for the most promiminent arcs)
the total integration time was 7200~sec.

%%%%%%%%%%%%%%%%%%%%%%%%%%%%%%%%%%%%%%%%%%%%%%%%
%%%%%%%%%%%%%%%%%%%%%%%%%%%%%%%%%%%%%%%%%%%%%%%%
%%%%%%%%%%%%%%%%%%%%%%%%%%%%%%%%%%%%%%%%%%%%%%%%
\section{Strong Lensing Analysis}
\label{sec:slens}
%%%%%%%%%%%%%%%%%%%%%%%%%%%%%%%%%%%%%%%%%%%%%%%%
%%%%%%%%%%%%%%%%%%%%%%%%%%%%%%%%%%%%%%%%%%%%%%%%
%%%%%%%%%%%%%%%%%%%%%%%%%%%%%%%%%%%%%%%%%%%%%%%%

%%%%%%%%%%%%%%%%%%%%%%%%%%%%%%%%%%%%%%%%%%%%%%%%
\subsection{Data}
%%%%%%%%%%%%%%%%%%%%%%%%%%%%%%%%%%%%%%%%%%%%%%%%

We obtain strong lensing constraints of the clusters by fitting the
positions of arcs. Figure~\ref{fig:img} presents the {\it Subaru}
Suprime-cam images of the central regions of the clusters. As shown in
\citet{hennawi08}, these clusters exhibit clear giant arcs. We identify
multiple images based on their colors, redshifts, and also iteratively
by building preliminary mass models of the clusters. For A1703, we
adopt the identification and spectroscopic redshifts of arcs reported
by \citet{limousin08} and \citet{richard09}.  The redshifts of giant
arcs of SDSS1531 were successfully measured with the {\it Gemini}
telescope. Although we could not measure the redshift of the lensed
blue images in SDSS1446 from our {\it Gemini} spectrum, the absence of
[OII] 3727{\AA} emission line (which should be detected for blue,
actively star-forming galaxies like this) and the brightness of the
arc in the $g$-band image suggests that the redshift should be in the
range $1.6<z<3.5$ (Hennawi et al. 2009, in preparation). We adopt this
redshift range in our mass modeling. Since the spectrum of the giant
arc of SDSS2111 was inconclusive, we conservatively assume its
redshift to be $z<3.5$ based on the $g-r$ color. Although some
additional possible multiply imaged systems are identified in some of
these clusters, we conservatively restrict our analysis to those image
systems where lensing hypothesis is completely unambiguous. 
We summarize the locations and redshifts of the multiply
imaged systems for strong lens modeling in Table~\ref{tab:strong}.   

%%%%%%%%%%%%%%%%%%%%%%%%%%%%%%%%%%%%%%%%%%%%%%%%
\subsection{Mass Modeling}
%%%%%%%%%%%%%%%%%%%%%%%%%%%%%%%%%%%%%%%%%%%%%%%%
We model the dark matter distribution in each clusters by the
NFW profile:
%%%%%%%%%%%%%%%%%%
\begin{equation}
\rho(r)=\frac{\rho_s}{(r/r_s)(1+r/r_s)^2},
\label{eq:nfw}
\end{equation}
%%%%%%%%%%%%%%%%%%
where $\rho_s$ is a characteristic density and $r_s$ is a scale radius.
Throughout the paper, we adopt the virial overdensity $\Delta_{\rm
  vir}(z)$ to compute the
virial mass $M_{\rm vir}$ and virial radius $r_{\rm vir}$:
%%%%%%%%%%%%%%%%%%
\begin{equation}
M_{\rm vir}=\frac{4}{3}\pi r_{\rm vir}^3\Delta_{\rm
  vir}(z)\bar{\rho}(z)=\int_0^{r_{\rm vir}}4\pi r^2\rho(r)dr,
\end{equation}
%%%%%%%%%%%%%%%%%%
where $\bar{\rho}(z)$ is the mean matter density of the universe at redshift
$z$. The nonlinear overdensity $\Delta_{\rm vir}(z)$ is calculated at
each $z$ in a standard way using the spherical collapse model. The
concentration parameter $c_{\rm vir}$ of the model is defined by  
%%%%%%%%%%%%%%%%%%
\begin{equation}
c_{\rm vir}=\frac{r_{\rm vir}}{r_s}.
\end{equation}
%%%%%%%%%%%%%%%%%%
Thus clusters with larger values of $c_{\rm vir}$ are more centrally
concentrated. $N$-body simulations based on the $\Lambda$CDM model
predict that concentration parameters depend on halo masses and
redshifts; for massive clusters, they are typically $\sim 5$ at $z=0$
and evolve with the redshift as $(1+z)^{-\alpha}$ with $\alpha\sim 0.7-1$ 
\citep{bullock01,wechsler02,zhao03,hennawi07,neto07,duffy08,maccio08}. 

To compute lensing properties, we first project the spherical NFW
profile (eq. [\ref{eq:nfw}]) onto a two-dimensional lens plane. Next
we include an ellipticity $e$ by replacing the radius
$r=\sqrt{x^2+y^2}$ in the projected mass density to
$\sqrt{(1-e)x^2+y^2/(1-e)}$. Then we allow an arbitrary position angle
$\theta_e$ by rotating the projected mass density. The lensing
deflection angle for our mass profile is computed using the method
described in \citet{schramm90}. Since the 
ellipticity is introduced in the surface mass density rather than the
lens potential,  our model does not suffer from unphysical mass
distributions (such as dumbell-like isodensity contours and negative
mass densities), which may be seen in the case of elliptical lens
potentials \citep[e.g.,][]{golse02}.

We also include member galaxies in our mass modeling. The
total mass distribution of each member galaxy is modeled by the
pseudo-\citet{jaffe83} model which has the three-dimensional radial
profile of $\rho\propto r^{-2}(r^2+r_{\rm cut}^2)^{-1}$, i.e., the
isothermal model with the truncation at $r_{\rm cut}$. We 
identify red member galaxies in the color-magnitude ($g-i$ versus
$r$) diagram and include $\sim 40$ brightest member galaxies for the
mass model of each cluster. We fix the locations, ellipticities, and
position angles of the members to those measured in the {\it Subaru}
Suprime-cam data.  Note that including the central galaxies is
important for accurate estimates of dark matter core masses, as their
effects on strong lensing can be significant
\citep[e.g.,][]{meneghetti03,hennawi07,wambsganss08}. 
On the other hand, the effect of other member galaxies is rather
minor, but they can be important in accurately reproducing positions
of some lensed images.  

To reduce the number of parameters, we adopt the exact scaling
relations for the velocity dispersions $\sigma$ and cutoff radii
$r_{\rm cut}$ of member galaxies following \citet{natarajan97}, who
constrained the mass distribution around cluster elliptical galaxies
using detailed strong and weak lensing models. Specifically, we assume
that they scale as $\sigma\propto L^{1/4}$ (Faber-Jackson relation) and 
$r_{\rm cut}\propto L^{1/2}$ (corresponding to a constant
mass-to-light ratio), where $L$ is a total $r$-band luminosity of the
galaxy. We apply this scaling relation to all member galaxies
including the BCG of the cluster.  To avoid unrealistically massive
member galaxies, we add a Gaussian prior to the normalization of the
scaling for the velocity dispersion, which is estimated using the
correlation with the magnitude \citep{bernardi03}, but we adopt a
conservative $20\%$ error so that a wide range of mass models are
allowed. We also estimate the normalization of the scaling for the
cutoff radius by that inferred from observations,  
$r_{\rm cut}=18(\sigma/200{\rm km\,s^{-1}})^2$~kpc (where we fix
$\sigma$ in this relation to the mean normalization of the Gaussian
prior on $\sigma$) \citep{natarajan02,natarajan09}, but again allow
$r_{\rm cut}$ to be Gaussian-distributed with a $50\%$ error which
crudely models the uncertainty of this relation in observations.  

Finding multiple images for a given source position requires intensive
search in the image plane. This is time-consuming, even if we adopt
adaptive-mesh refinement scheme to improve grid resolutions only near
the critical curves, preventing us from exploring a large parameter
space in a reasonable time scale. Instead, in this paper we evaluate
$\chi^2$ in the source plane to speed up \citep[e.g.,][]{kochanek91}:
%%%%%%%%%%%%%%%%%%
\begin{equation}
\chi^2_{\rm src}=\sum_i\frac{\left|\mu_i(\mathbf{u}_i-\mathbf{u}_{\rm
    src})\right|^2}{\sigma_{\rm pos}^2}+{\rm prior},
\end{equation}
%%%%%%%%%%%%%%%%%%
where $\mu_i$ is the magnification tensor for observed $i$-th image,
  $\mathbf{u}_i$ are the source position computed from the position
of the $i$-th image, and $\mathbf{u}_{\rm src}$ is the best-fit
source position which can be found analytically 
 \citep[see][]{keeton01}. Since the magnification tensor is nothing
 but a mapping from the source plane to the image plane,
 $|\mu_i(\mathbf{u}_i-\mathbf{u}_{\rm src})|$ approximates the
 distance between observed and predicted image positions in the image
plane; indeed, we confirmed that this source plane $\chi^2$ is
reasonably accurate at around our best-fit models, compared with the
standard image plane $\chi^2$. Note that this approach automatically
ignores any extra images.  An important parameter here is the
positional uncertainty in the image plane, $\sigma_{\rm pos}$, which
is directly related with errors on best-fit model parameters. While
the measurement error was sufficiently small, comparable to the pixel
scale of the image ($\sim 0\farcs2$), it has been known that
multiple images can be fitted only with $\sim 1''$ accuracy for some
massive lensing clusters \citep[e.g.,][]{broadhurst05b}, presumably
because of complex nature of cluster mass distributions. Thus, in this
paper we assume the positional error of $1\farcs2$ (image plane) for
all multiple images.  

We explore the multi-dimensional $\chi^2$ surface using a Markov chain
Monte Carlo (MCMC) approach. We adopt a standard Metropolis-Hastings
sampling with the multivariate-Gaussian as a proposal distribution. 
During the sampling, the range of the virial mass is restricted to
$1.4\times 10^{14}M_\odot< M_{\rm vir} < 7\times 10^{15}M_\odot$, and
that of the concentration parameter to $c_{\rm vir}<40$. We then
derive constraints on parameters by projecting the likelihood
distributions to the parameter space. Mass modeling is performed using
the software {\it glafic} (M. Oguri, in preparation). 

%%%%%%%%%%%%%%%%%%%%%%%%%%%%%%%%%%%%%%%%%%%%%%%%
\subsection{Result of Strong Lens Modeling}
\label{sec:slresult}
%%%%%%%%%%%%%%%%%%%%%%%%%%%%%%%%%%%%%%%%%%%%%%%%

\begin{figure}
\epsscale{0.48}
\plotone{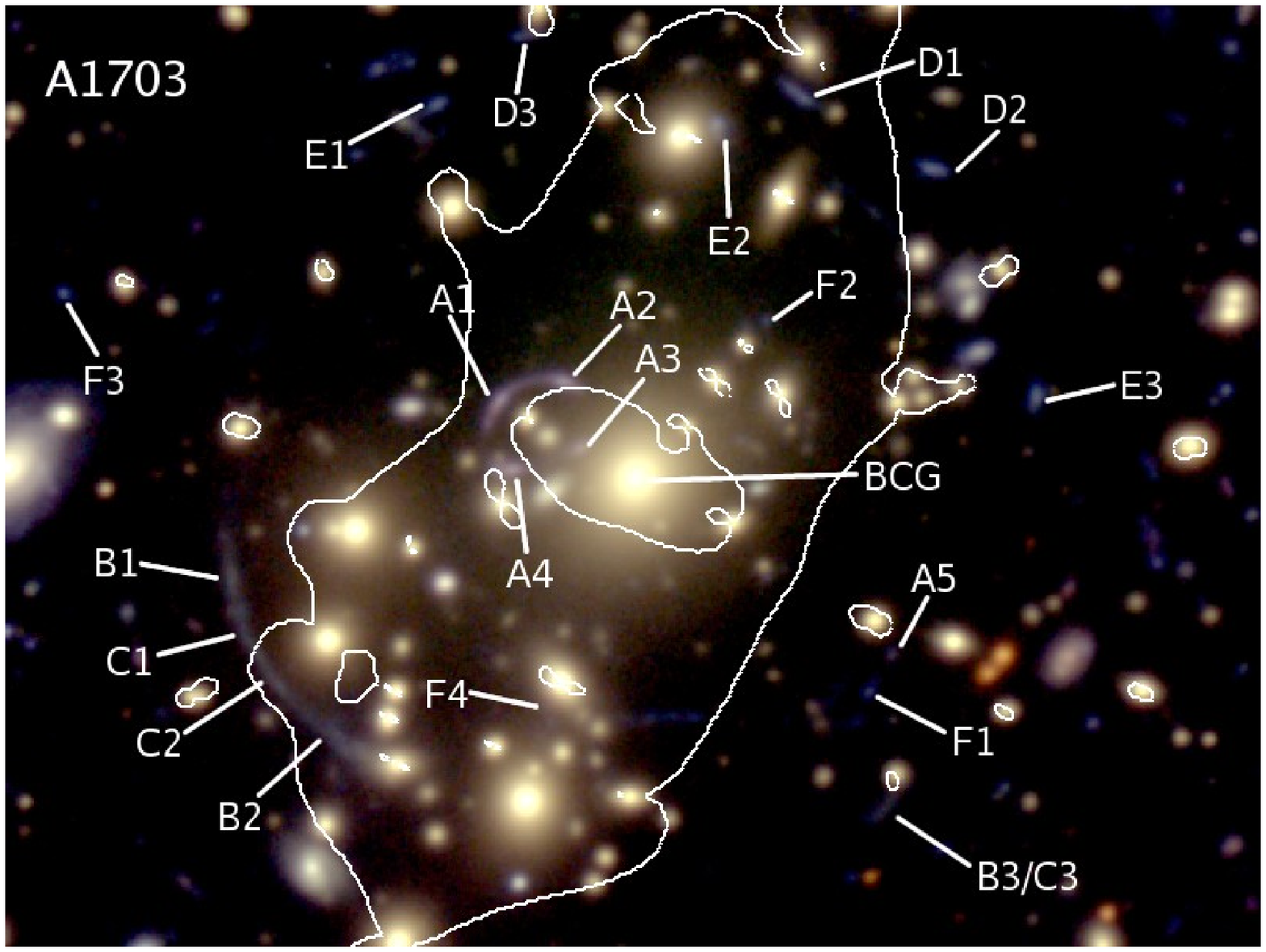}
\plotone{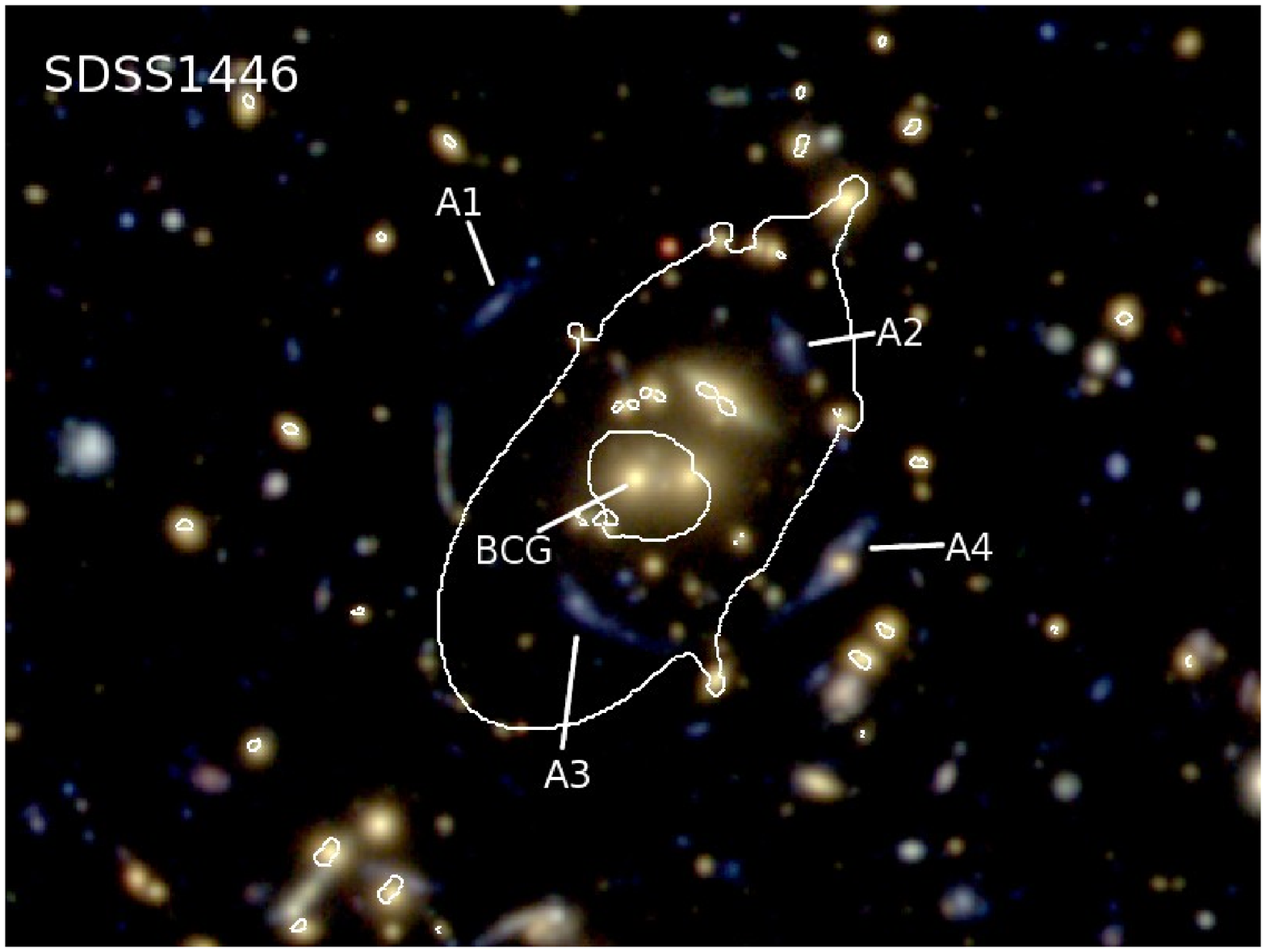} \\
\plotone{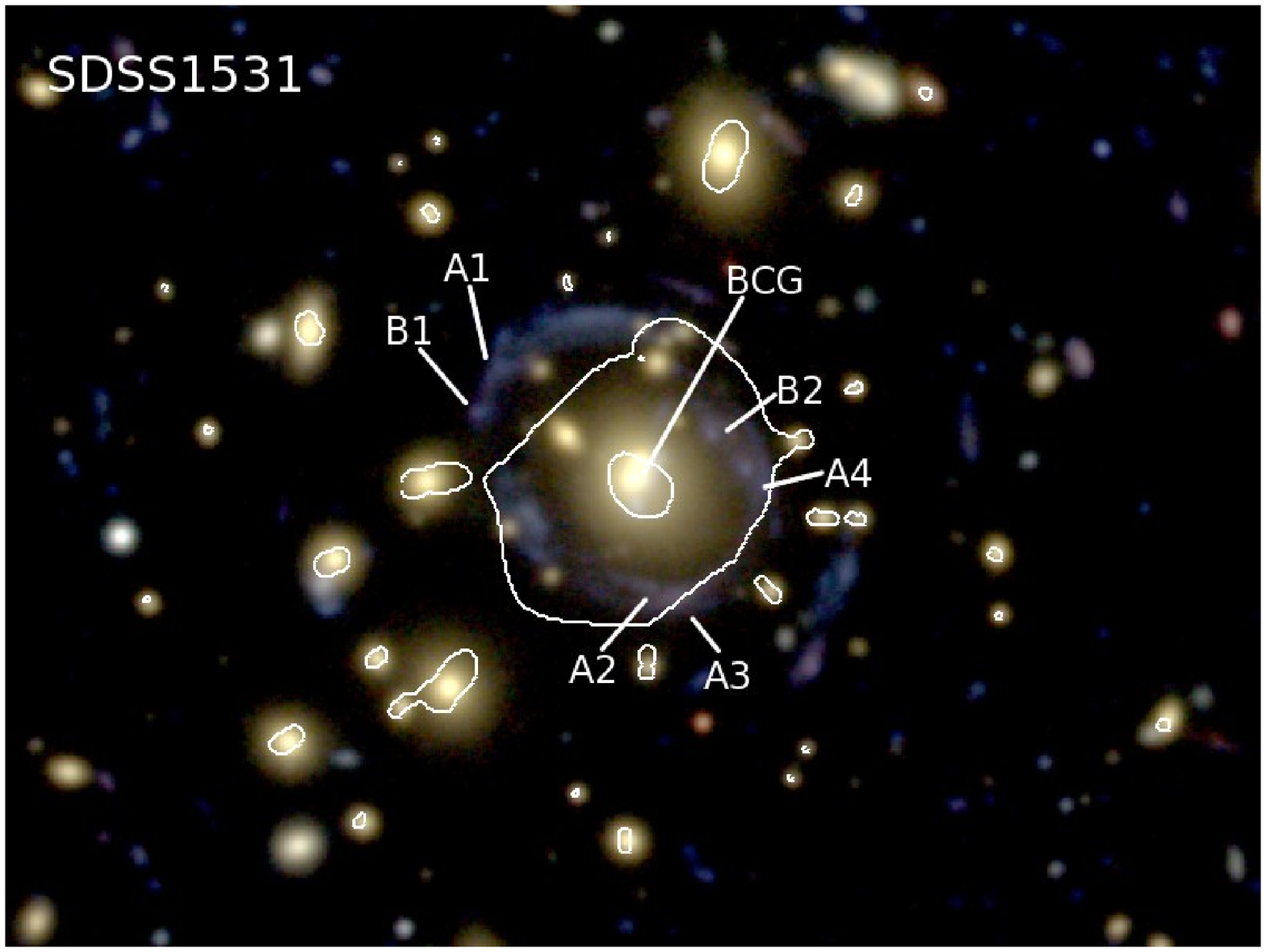}
\plotone{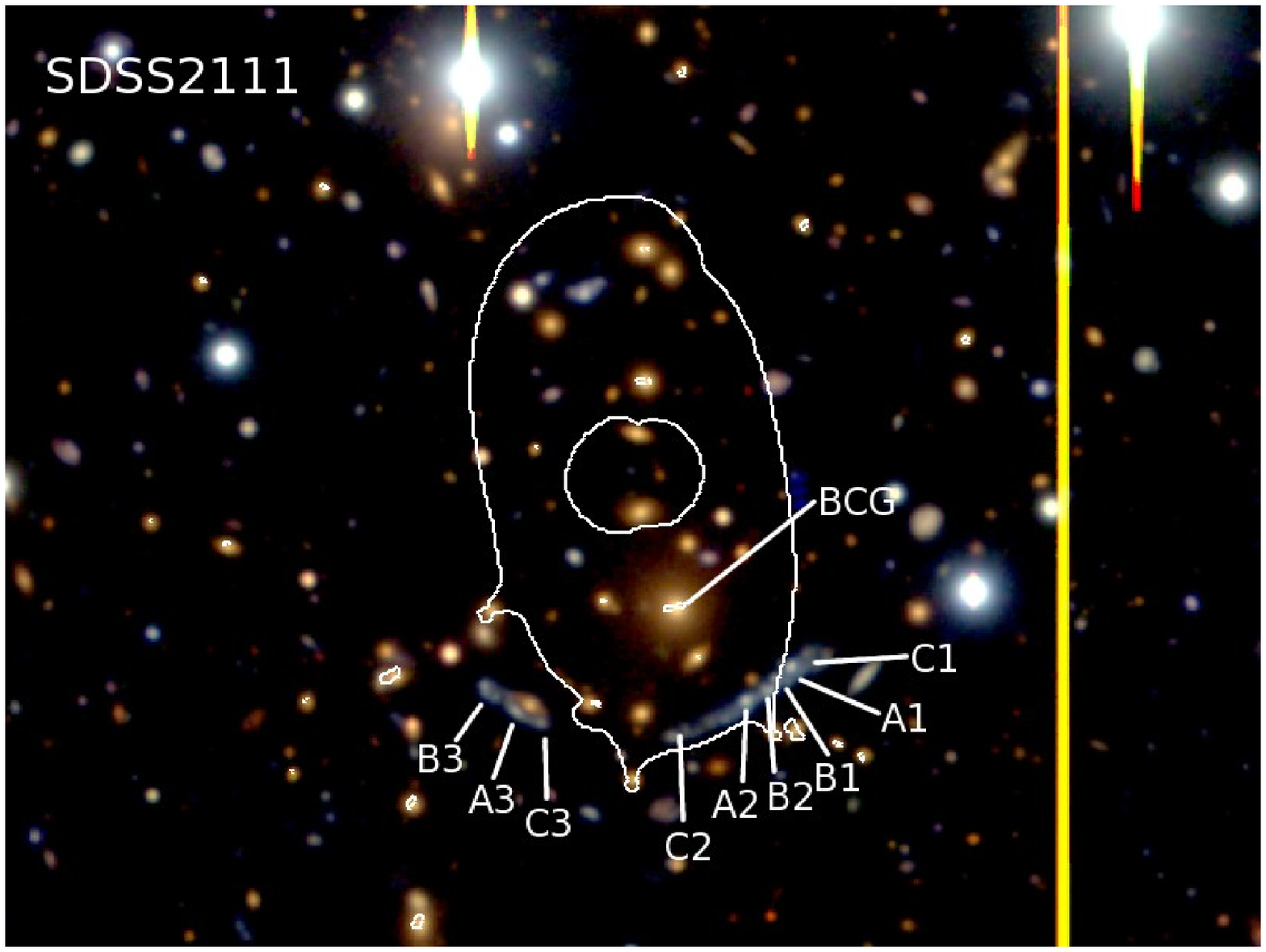}
\caption{{\it Subaru} Suprime-cam images of 4 clusters we study in
  this paper. North is up and East is left. The size of each image is
  $107'' \times 80''$. Multiple images we adopt for strong lens
  modeling are named by an alphabet and a number (e.g., A1; see also
  Table~\ref{tab:strong}). Critical curves of our best-fit models at
  the redshifts of multiple images ($z=2.627$ for A1703) are drawn by
  solid lines (see \S\ref{sec:slresult}). The brightest cluster galaxy
  for each cluster is marked by ``BCG''. The J2000.0 
  coordinates of the BCGs are  13:15:05.24+51:49:02.7 (A1703),
  14:46:34.03+30:32:58.7 (SDSS1446), 15:31:10.63+34:14:24.9
  (SDSS1531), and 21:11:19.36$-$01:14:22.9 (SDSS2111).
\label{fig:img}}  
\end{figure}

Results of our strong lens modeling are summarized in
Table~\ref{tab:slmodel}, and the critical curves of the best-fit
models are displayed in Figure~\ref{fig:img}.  First, we find that our
best-fit models successfully reproduce the multiple images with
reduced chi-squares of $\chi^2/{\rm dof}\lesssim 1$, which suggests
that our assumption of the positional error of $1\farcs 2$ was
reasonable. Second, strong lensing constrains the center of the dark
halo component quite well with a typical error of an arcsecond, and
the best-fit dark halo center is close to the position of the BCG. An
exception is SDSS2111 for which the offset of the center of the dark
halo from the BCG is large, $> 10''$ (see below). Third, it is
found that the virial mass $M_{\rm vir}$ and concentration parameter
$c_{\rm vir}$ are not constrained very well. In fact this is expected,
because multiple images of a source redshift mainly constrain the mass
enclosed by the arcs, which is sensitive both to $M_{\rm vir}$ and
$c_{\rm vir}$. Fourth, we find that relatively large  ellipticities of
$e\sim 0.4$ are required to fit the images. An exception is SDSS1531
for which the mass distribution appears to be rather circular, $e\sim
0.1$, although an ellipticity as high as $\sim 0.5$ is allowed. These
large ellipticities are in fact broadly consistent with $\Lambda$CDM
predictions \citep[e.g.,][]{oguri09}. 

Our best-fit mass model of A1703 from strong lensing (see also
Appendix~\ref{sec:gnfw}) appears to be in reasonable agreement with
earlier work \citep{limousin08,saha09,richard09}. For example,
best-fit ellipticity and position angle of dark  halo component,
$e=0.36$ and $\theta_e=-24.1$, agree well with the result of \citet{richard09},
the ellipticity in the lens potential of $e_{\rm pot}=0.23$ and
$\theta_e=-26.0$. \citep[see, e.g.,][ for the relation between $e_{\rm
 pot}$ and $e$]{golse02}.  Moreover, we find that the structure of
the critical curve of A1703  is consistent with those found in the
earlier work. It is worth noting that the central ring (A1--A4)
represents a rare example of lensing by a hyperbolic umbilic
catastrophe \citep[e.g.,][]{blandford86}. See \citet{orban09} for more
comprehensive discussions about the central ring as a hyperbolic
umbilic catastrophe. 

One of the most important quantities that are well constrained by
strong lensing is the Einstein radius, $\theta_{\rm Ein}$. We obtain
constraints on $\theta_{\rm Ein}$ at the redshifts of the arcs 
from our strong lens modeling as follows. For all the MCMC samples we 
estimate the Einstein radii of the dark halo (NFW) components by
forcing $e=0$ and calculating the radii of the outer critical curves
predicted by the dark halo components alone.\footnote{We compared
  radial profiles of dark halo components with $e=0$ with
  azimuthally averaged radial profiles of elliptical dark halo
  components, and found that they agree with each other at
  $\lesssim$3\% level, which is negligibly small compared with
  statistical errors. Thus the procedure adopted in this paper does
  not involve any significant systematics.}  Thus the contribution of
cluster member galaxies including the central galaxy is removed in
our estimate of $\theta_{\rm Ein}$, suggesting that the value could in
fact be smaller than observed distances of lensed images from the
center. We emphasize that this quantity serves as a useful strong lens
constraint when combined with weak lensing results, particularly
because we are interested in the radial profile of the dark matter
component, rather than the total matter including stars which dominate
at the very center of clusters (see
\S\ref{sec:s+w}).\footnote{Strictly speaking, the Einstein radius
  derived in this way actually constrains the mass distribution of
  dark matter plus hot gas, rather than the dark matter distribution
  alone.} The derived constraints on $\theta_{\rm Ein}$ are shown in
Table~\ref{tab:slein}. Since multiple strong lens systems with
different redshifts are available for A1703, for this particular
cluster we derive Einstein radii at two different source redshifts,
$z=0.8889$ (the redshift of arc A1$-$A5) and $2.627$ (the redshift of
arc B1$-$B3 and C1$-$C3). We regard these two Einstein radii as
statistically independent, which is reasonable given the large
difference of $\theta_{\rm Ein}$ between these two redshifts. 

Here we briefly discuss the mass model of SDSS2111, which appears to
exhibit the large offset between the lens potential center and the
location of the BCG (see Figure~\ref{fig:img}; the best-fit lens
potential center corresponds to the center of the inner critical
curve). We find that no acceptable lens model has the offset smaller
than $\sim 5''$. The offset is required because the center of
curvature of the long bright arc is clearly displaced from the
BCG. However, as seen in Table~\ref{tab:slmodel}, the current data do
not constrain the center of the dark halo (NFW) component very well,
which is the main reason that $\theta_{\rm Ein}$ involve relatively
large errors.  Thus it is important to improve the strong lens
constraints by adding more multiply imaged systems. 

%%%%%%%%%%%%%%%%%%%%%%%%%%%%%%%%%%%%%%%%%%%%%%%%
%%%%%%%%%%%%%%%%%%%%%%%%%%%%%%%%%%%%%%%%%%%%%%%%
%%%%%%%%%%%%%%%%%%%%%%%%%%%%%%%%%%%%%%%%%%%%%%%%
\section{Weak Lensing Analysis}
\label{sec:wlens}
%%%%%%%%%%%%%%%%%%%%%%%%%%%%%%%%%%%%%%%%%%%%%%%%
%%%%%%%%%%%%%%%%%%%%%%%%%%%%%%%%%%%%%%%%%%%%%%%%
%%%%%%%%%%%%%%%%%%%%%%%%%%%%%%%%%%%%%%%%%%%%%%%%

%%%%%%%%%%%%%%%%%%%%%%%%%%%%%%%%%%%%%%%%%%%%%%%%
\subsection{Distortion Measurements}
%%%%%%%%%%%%%%%%%%%%%%%%%%%%%%%%%%%%%%%%%%%%%%%%

\begin{figure}
\epsscale{0.55}
\plotone{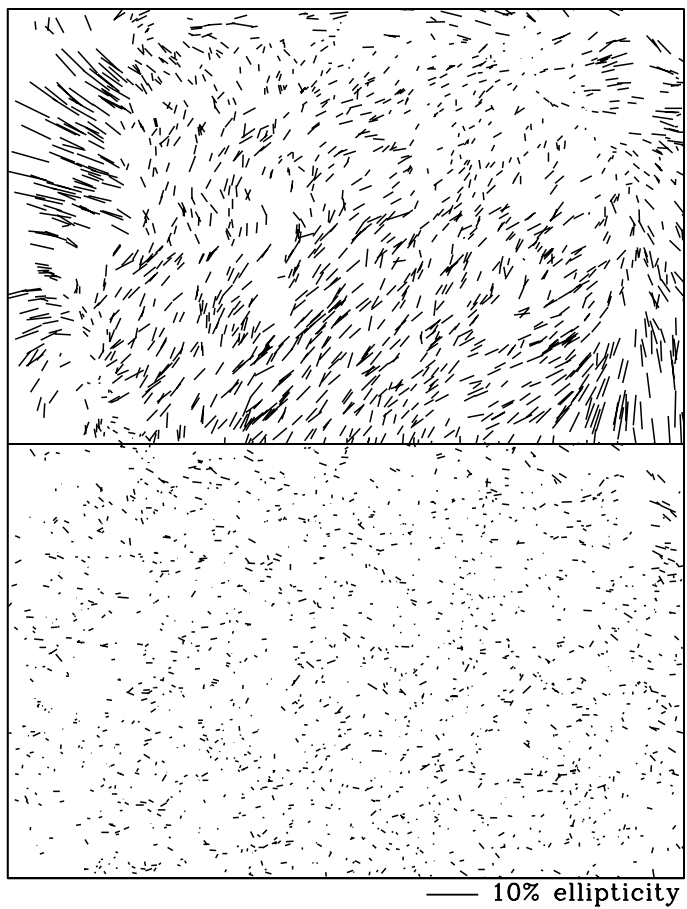}
\caption{PSF anisotropy maps of the filed of SDSS2111 measured from
  template stars. The upper panel shows the raw ellipticities of the
  stars, whereas the lower panel is stellar ellipticities after the
  PSF correction is applied. The length and orientation of each line
  segment indicates the ellipticity and position angle of each
  template star. The size of each panel is $\sim 35\farcm5\times
  26\farcm5$, i.e., the entire field of the mosaic image is shown
  here. Table~\ref{tab:wl} lists ellipticities before and after the
  PSF correction for the fields of all the clusters. 
\label{fig:ell2d}} 
\end{figure}

We derive weak lensing shear signals following the algorithm outlined
in \citet[][hereafter KSB]{kaiser95}. First, for each cluster we
reprocess {\it Subaru} Suprime-cam $r$-band images through
IMCAT\footnote{http://www.ifa.hawaii.edu/\~{}kaiser/imcat/} to 
compute shapes of each objects. In order to compute shapes accurately,
we iteratively refine the centroid of each object. During the process
objects with offsets in each iteration larger than 3~pixel are
removed. The ellipticities $\mathbf{e}$ of objects are measured from the
weighted quadruple moments of the surface brightnesses. We select
stars for Point Spread Function (PSF) correction  in a standard way by
identifying the stellar branch in the magnitude versus half light radius
$r_h$ plane. We apply an additional cut, the signal-to-noise ratio
$\nu>15$ and the ellipticity $|\mathbf{e}|<0.1$, to construct a final
sample of template stars for the PSF correction.  We correct
ellipticities of objects using the sample of stars as reference 
%%%%%%%%%%%%%%%%%%
\begin{equation}
\mathbf{e}_{\rm cor}=\mathbf{e}-P_{\rm sm}\mathbf{q}^*,
\end{equation}
%%%%%%%%%%%%%%%%%%
where $P_{\rm sm}$ is the smear polarizability tensor and
$\mathbf{q}^*=(P_{\rm sm}^*)^{-1}\mathbf{e}^*$ is the PSF anisotropy
kernel estimated from the template stars (hereafter the superscript
$^*$ denotes quantities measured for template stars). Following
\citet{okabe08} and \citet{umetsu08}, we obtain a smooth map of
$\mathbf{e}^*$ by first dividing the whole image into $5\times4$
chunks and fit $\mathbf{e}^*$ in each chunk using second-order
bi-polynomials with iterative 3$\sigma$-clipping rejections. We show
ellipticities of the template stars before and after the PSF
correction in Figure~\ref{fig:ell2d} and Table~\ref{tab:wl}. The
Figure demonstrates that our procedure successfully reduces PSF
anisotropies over the entire field. Residual stellar ellipticities
after the PSF correction are comparable to those in other weak lensing
studies using {\it Subaru} Suprime-cam images
\citep{hamana03,miyazaki07,okabe08,umetsu08}.  

Next we correct the isotropic smearing effect to estimate weak lensing
shear. The pre-seeing shear polarizability $P_g$ relates observed
galaxy ellipticities with their true values, and is calculated as 
%%%%%%%%%%%%%%%%%%
\begin{equation}
P_g=P_{\rm sh}-P_{\rm sm}(P_{\rm sm}^*)^{-1}P_{\rm sh}^*,
\end{equation}
%%%%%%%%%%%%%%%%%%
where $P_{\rm sh}$ is the shear polarizability tensor. Since the
measurement of $P_g$ for each galaxy is extremely noisy, we
approximate $P_g$ as $P_g^{\rm s}$ times an identity matrix with 
\citep{erben01,hetterscheidt07}: 
%%%%%%%%%%%%%%%%%%
\begin{equation}
P_g^{\rm s}=\frac{1}{2}\left\{{\rm tr}(P_{\rm sh})-\frac{{\rm tr}(P_{\rm sh}^*)}{{\rm tr}(P_{\rm sm}^*)}{\rm tr}(P_{\rm sm})\right\}.
\end{equation}
%%%%%%%%%%%%%%%%%%
Here we use a smooth map of ${\rm tr}(P_{\rm sh}^*)/{\rm tr}(P_{\rm sm}^*)$ 
constructed by fitting them as a function of the position, with a
fifth-order polynomial over the entire template stars. We then derive
the reduced shear $\mathbf{g}=\mbox{\boldmath{$\gamma$}}/(1-\kappa)$
for each object as 
%%%%%%%%%%%%%%%%%%
\begin{equation}
\mathbf{g}=f_{\rm cal}\frac{\mathbf{e}_{\rm cor}}{P_g^{\rm s}}.
\end{equation}
%%%%%%%%%%%%%%%%%%
We use raw, unsmoothed $P_g^{\rm s}$ because the smoothing does not
necessarily improve the shear estimate \citep[e.g.,][]{erben01}. We
include the calibration correction factor $f_{\rm cal}=1/0.88$ since
the KSB algorithm with the scalar correction scheme for $P_g$ has been 
known to underestimate shears by $\sim 10-15\%$
\citep{erben01,heymans06,massey07}. We confirmed that our shear
estimate procedure, together with the source galaxy selection
described in \S\ref{sec:wlsrc}, recovers lensing shears in the
simulated {\it Subaru} Suprime-cam images of the Shear Testing
Programme 2 \citep[STEP2;][]{massey07} quite well, with typically a
few percent of the shear-calibration bias (parameter $m$) and an order
of magnitude smaller residual shear offset (parameter $c$).

%%%%%%%%%%%%%%%%%%%%%%%%%%%%%%%%%%%%%%%%%%%%%%%%
\subsection{Source Galaxy Population}
\label{sec:wlsrc}
%%%%%%%%%%%%%%%%%%%%%%%%%%%%%%%%%%%%%%%%%%%%%%%%

The background galaxy selection is crucial for weak lensing analysis. 
We choose following galaxies for weak lensing analysis; (1) no close
companions at distances within 10~pixels, (2) the half light radius
$r_h$ larger than $1.2\overline{r_h^*}$, where $\overline{r_h^*}$ is
the median of the half light radii of template stars used for the PSF
correction, (3) the half-light radius $r_h$ smaller than 10~pixels,
(4) the signal-to-noise ratio $\nu>7$,  (5) the pre-seeing shear
polarizability $P_g^{\rm  s}>0.05$, and (6) the shear value
$|\mbox{\boldmath{$\gamma$}}|<2.0$.  In addition, we apply magnitude
and color cuts, as we will discuss below.    

An appropriate color selection is important to minimize the dilution
effect by cluster member galaxies \citep[e.g.,][]{medezinski07}.
One way to avoid contaminations from cluster members is to use
galaxies redder than the red sequence of the cluster or bluest
galaxies. In this paper we adopt blue galaxies, because our clusters  
are located at moderately high redshifts ($z\sim 0.3-0.7$) and hence
very few red background galaxies are available. Specifically, we use
galaxies with $g-i<1.0$ (measured in aperture magnitudes) for our weak
lensing analysis. This is much bluer than red-sequence galaxies
($g-i\sim 1.7-2.8$ for $r\sim 24$), suggesting that the dilution
effect is not significant for our source galaxy samples. 

We now estimate the depths of our galaxy sample for weak lensing
analysis. A key quantity here is the mean lensing depth averaged
over the population of source galaxies:
%%%%%%%%%%%%%%%%%%
\begin{equation}
\left\langle\frac{d_{ls}}{d_{os}}\right\rangle=\int dz
\frac{dp_{\rm wl}}{dz}\frac{d_{ls}}{d_{os}},
\label{eq:lensdepth}
\end{equation}
%%%%%%%%%%%%%%%%%%
with $d_{ls}$ and $d_{os}$ being angular diameter distances from
the lens to the source and from the observer to the source,
respectively. The PDF $dp_{\rm wl}/dz$ denotes the redshift
distribution of the source galaxy sample, which is particularly
important in our study because of relatively high redshifts of the
clusters. 

\begin{figure}
\epsscale{0.55}
\plotone{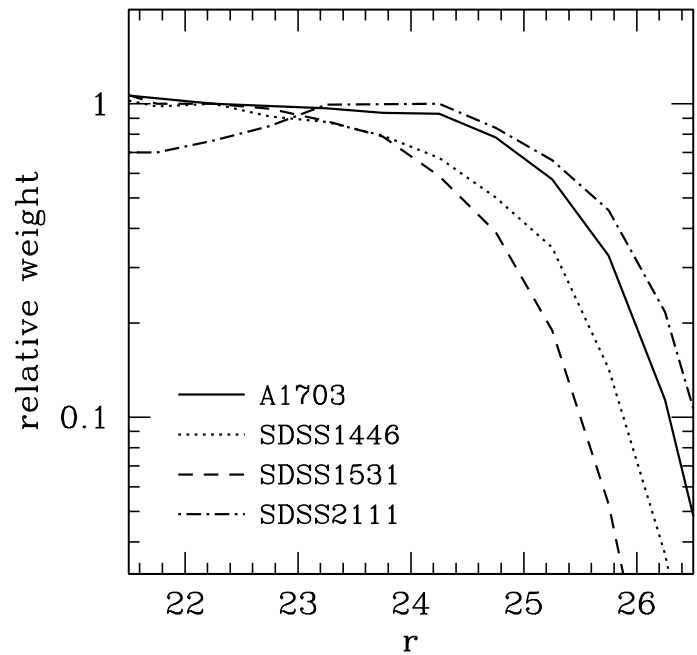}
\caption{The relative weight $w(r)$ of our background galaxy samples
  as a function of total $r$-band magnitudes, derived from
  the number ratio of source galaxies for weak lensing analysis and
  galaxies in the CFHTLS \citep{ilbert06}. The color cut $g-i<1.0$ is
  applied to both galaxy samples. The weights are normalized by 
  the maximum weight at $r>22$.  
\label{fig:wei}} 
\end{figure}

To estimate $dp_{\rm wl}/dz$, we adopt a photometric
redshift sample of galaxies in the Canada-France Hawaii Telescope
Legacy Survey (CFHTLS) presented by \citet{ilbert06}. The idea is that
we apply our color and magnitude cuts to the CFHTLS galaxy sample,
which is well calibrated by the large amount of photometric and
spectroscopic data, to infer the redshift distribution in our {\it
  Subaru} galaxy sample. First, we  compare the number counts of
galaxies with $g-i<1.0$ between the CFHTLS and our source galaxy
samples. Figure~\ref{fig:wei} shows the relative weight as a function
of total $r$-band magnitudes, $w(r)\propto n_{\rm Subaru}(r)/n_{\rm CFHTLS}(r)$,
i.e., the number ratio of the source galaxies and CFHTLS
galaxies. Sharp declines at $r\sim 25$ imply that the effect of our 
background galaxy selections is significant beyond this magnitude. We
therefore use galaxies with total $r$-band magnitudes 
$22<r<r_{\rm max}$ for our weak lensing analysis, where $r_{\rm max}$
is chosen so that the relative weight becomes approximately 0.3 (see
Table~\ref{tab:zs} for exact values). Next we estimate the probability
distribution functions (PDFs) of redshifts for our background samples
$dp_{\rm wl}/dz$ as 
%%%%%%%%%%%%%%%%%%
\begin{equation}
\frac{dp_{\rm wl}}{dz}=\frac{\sum w(r)dp/dz(r)}{\sum w(r)},
\end{equation}
%%%%%%%%%%%%%%%%%%
where the summations run over all CFHTLS galaxies with $g-i<1.0$ and
$22<r<r_{\rm max}$, and $dp/dz(r)$ is the PDFs of photometric
redshifts for individual CFHTLS galaxies \citep{ilbert06}. 
With this PDF, we estimate the lensing depths 
$\left\langle d_{ls}/d_{os}\right\rangle$ from equation 
(\ref{eq:lensdepth}). 

Table~\ref{tab:zs} summarizes the mean depths of
our source galaxy sample. We also compute the effective source
redshift $z_d$ that is defined by the redshift at which the distance
ratio $d_{ls}/d_{os}$ becomes equal to the mean distance ratio $\langle
d_{ls}/d_{os}\rangle$; in most situations of weak lensing analyses 
we can simply assume that all the source galaxies lie at $z=z_d$. 
However, since the reduced shear
$\mathbf{g}=\mbox{\boldmath{$\gamma$}}/(1-\kappa)$ is not exactly
proportional to $d_{ls}/d_{os}$, adopting a common redshift for all
background galaxies in fact underpredicts values of $\mathbf{g}$
\citep{seitz97,hoekstra00}. Thus, in computing expected $\mathbf{g}$
for a given mass model we fully take account of the redshift
distribution of background galaxies $dp_{\rm wl}/dz$ instead of
adopting the effective source redshift $z_d$. 

The resulting numbers of background galaxies are $\sim 9000-10000$
(surface number density of $n_g\sim 10\,{\rm  arcmin}^{-2}$), except
SDSS1531 for which the number of source galaxies is much smaller,
$\sim 6000$ ($n_g\sim 7\,{\rm arcmin}^{-2}$), because of the worse
seeing size of the image. The reason for the relatively small number
density compared with the expected  number density of $n_g\sim
30-40\,{\rm  arcmin}^{-2}$ for {\it Subaru} weak lensing
\citep[e.g.,][]{miyazaki07} is our color cut to eliminate the dilution
effect. In fact, the number density is comparable to other {\it
  Subaru} cluster weak lensing studies which adopted similar color
cuts \citep[e.g.,][]{okabe08}.   

%%%%%%%%%%%%%%%%%%%%%%%%%%%%%%%%%%%%%%%%%%%%%%%%
\subsection{Two-Dimensional Mass Distributions}
%%%%%%%%%%%%%%%%%%%%%%%%%%%%%%%%%%%%%%%%%%%%%%%%

Before studying radial profiles, we check two-dimensional mass
(convergence) distributions reconstructed from weak lensing shears
using an inversion algorithm of \citet{kaiser93}. Here we assume weak
lensing limit ($\kappa\ll1$, i.e.,
$\mathbf{g}\approx\mbox{\boldmath{$\gamma$}}$), which suggests that
the reconstructed $\kappa$ will be slightly overestimated in the cores
of clusters. We use these two-dimensional mass maps only to study the
morphology of the cluster; we constrain cluster mass distributions
using radial profiles of tangential shears (see \S\ref{sec:rsh}). We
also derive luminosity density maps of red-sequence galaxies for
comparison.  

\begin{figure}
\epsscale{0.48}
\plotone{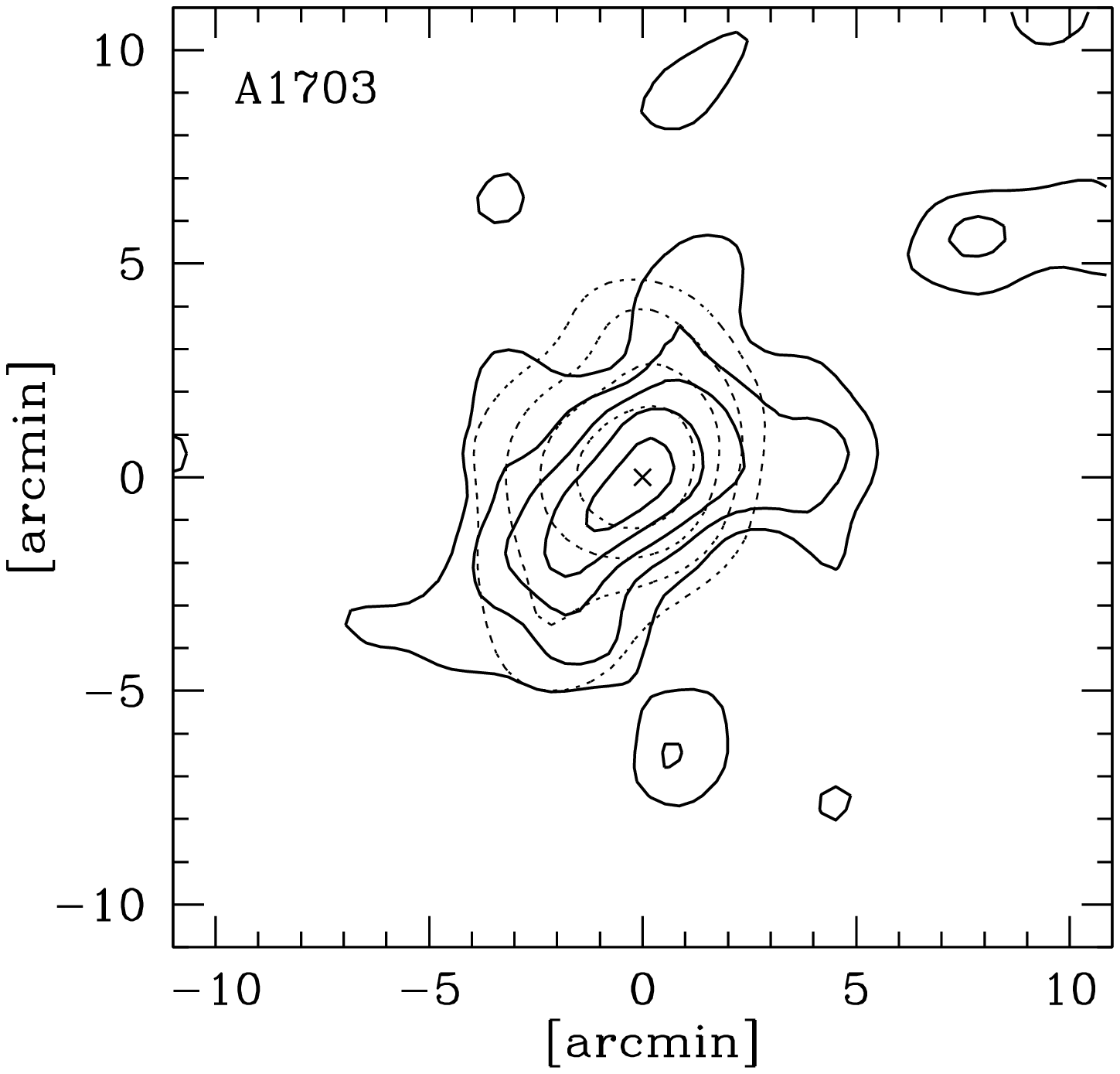}
\plotone{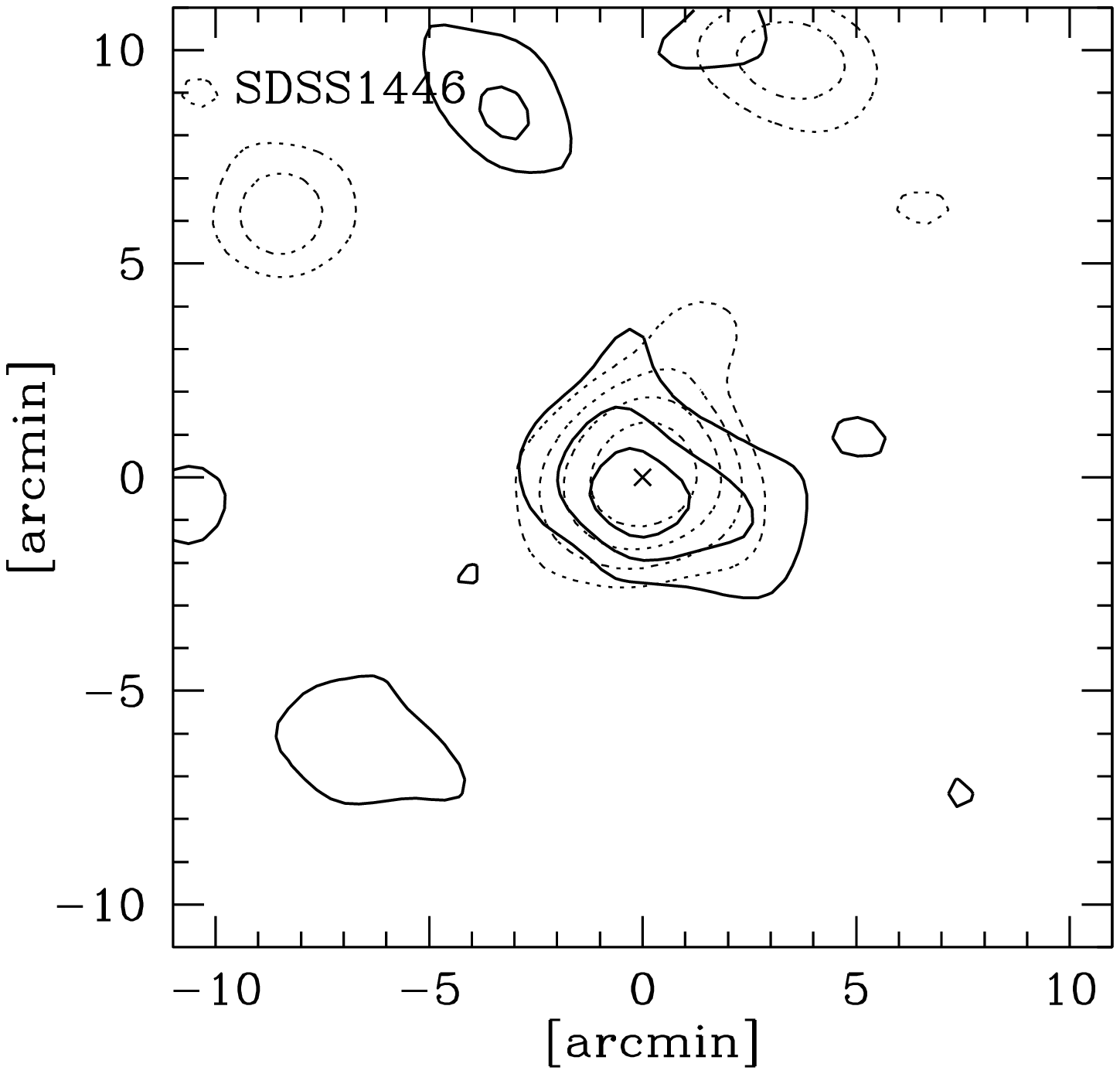} \\
\plotone{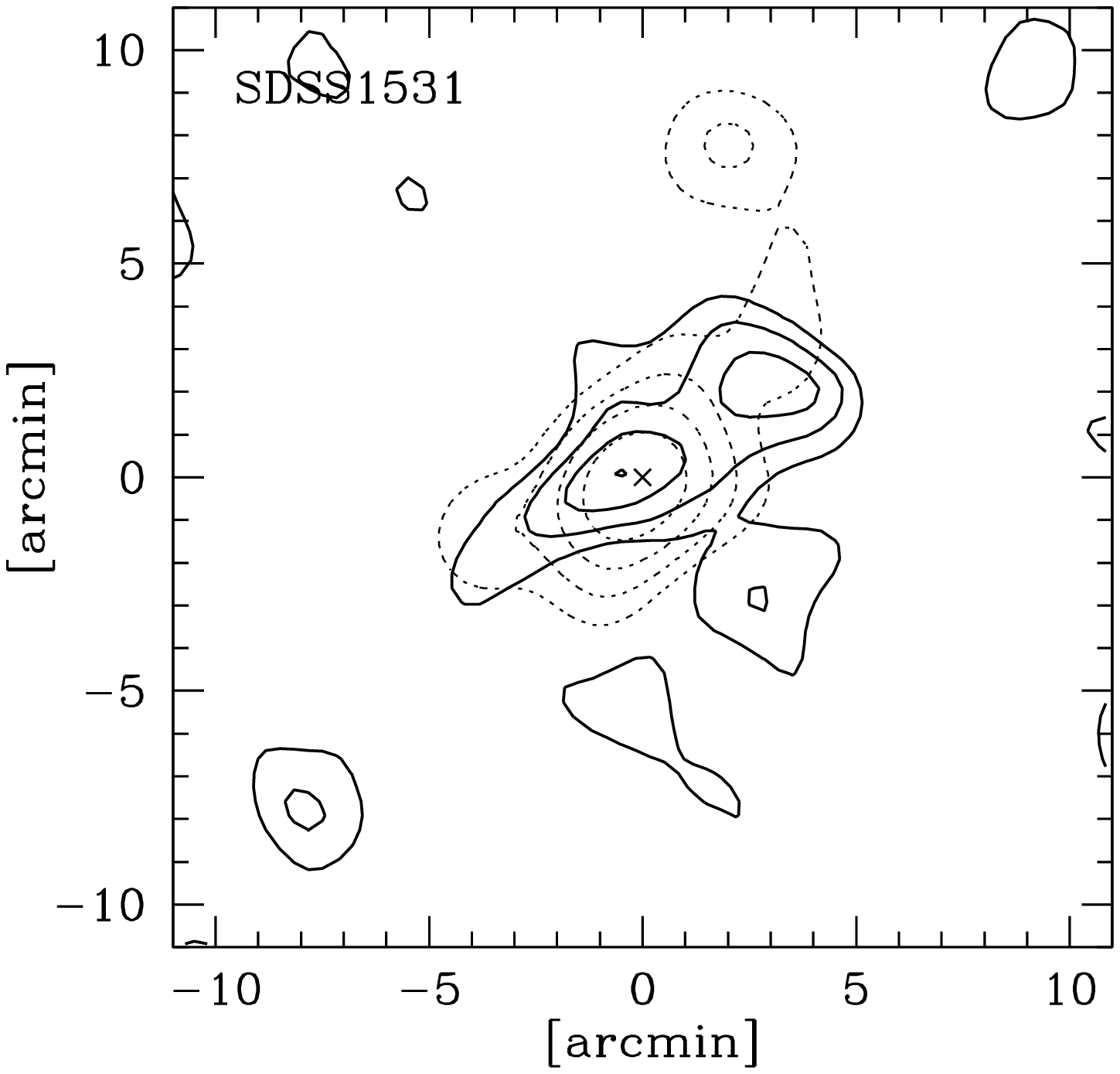}
\plotone{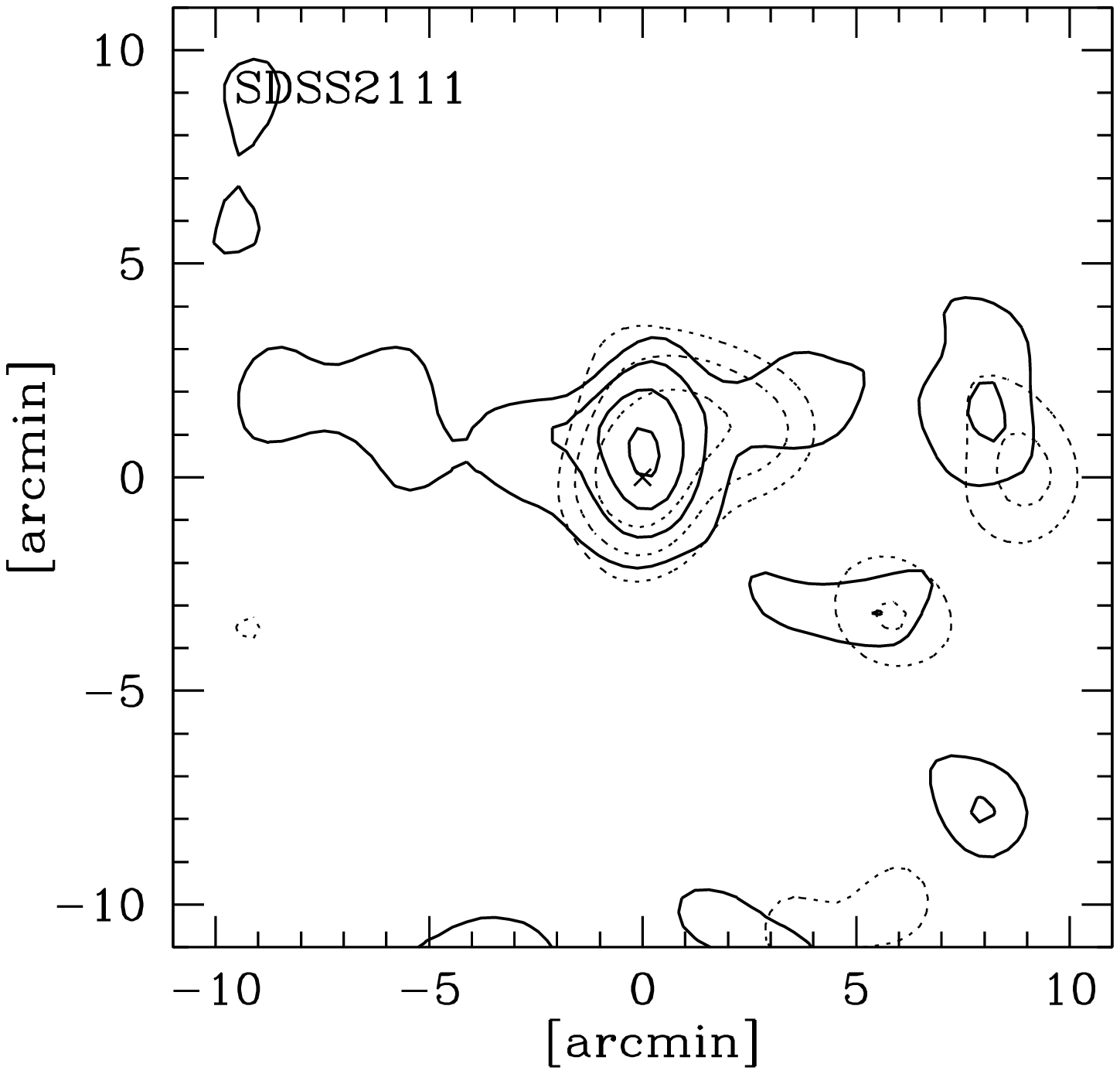}
\caption{Two-dimensional mass maps ({\it solid}) reconstructed from
  weak lensing shear measurements. North is up and East is left.
  Contours are drawn with spacing of $\Delta\kappa=0.04$, starting
  from $\kappa=0.04$. Note that the reconstructed $\kappa$ are
  slightly overestimated in the cores of clusters because we assumed
  the weak lensig limit,
  $\mathbf{g}\approx\mbox{\boldmath{$\gamma$}}$. The 1$\sigma$ noise 
  level of each reconstructed mass map is $\sigma_\kappa\sim
  0.03$. Crosses indicate the locations of the BCGs. Contours of
  constant luminosity densities of red-sequence galaxies ({\it
  dotted}) are also shown for comparison. Both mass and luminosity
  maps are Gaussian-smoothed with $\sigma=60''$.  
\label{fig:twod_kappa}} 
\end{figure}

We show Gaussian-smoothed ($\sigma=60''$) maps of all four clusters 
in Figure~\ref{fig:twod_kappa}. In all the clusters, weak lensing
signals are clearly detected. In addition, we find that the
reconstructed mass distributions well trace the spatial distributions
of red-sequence galaxies, with roughly similar mass and light
centroids. The clusters exhibit deviations from circular mass
distributions, in particular for A1703. The umidordal structure with
aligned position angles of weak lensing, the light distribution, and
strong lensing (see the critical curve shown in Figure~\ref{fig:img})
implies that the cluster A1703 is quite relaxed. For SDSS1531, we find
a probable secondary peak at the $\sim 4'$ northwest of the cluster
center, implying that the cluster may not yet be relaxed. The position
angles of mass distributions from strong and weak lensing appear to
agree. Several subpeaks and non-regular morphology of SDSS2111 suggest
that this cluster is not relaxed either. However, SDSS2111 again shows
agreement between weak and strong lensing in that the isodensity
contours appear to be elongated in the north-south direction. In
addition, the centroid of the mass distribution is slightly offset
North of the BCG, which is consistent with the result of strong lens
modeling. It is important to obtain X-ray data for these clusters to
explore their dynamical states further. 

%%%%%%%%%%%%%%%%%%%%%%%%%%%%%%%%%%%%%%%%%%%%%%%%
\subsection{Tangential Shear Profiles}
\label{sec:rsh}
%%%%%%%%%%%%%%%%%%%%%%%%%%%%%%%%%%%%%%%%%%%%%%%%

For each cluster we derive an azimuthally averaged one-dimensional
shear profile to obtain constraints on the cluster mass profiles.
The center of each cluster is fixed to the center of the dark halo
component determined from strong lens modeling (see
Table~\ref{tab:slmodel}). For a given center, we can calculate the
tangential shear $g_+$ and the $45^\circ$ rotated component $g_\times$
from the reduced shear $\mathbf{g}=(g_1,\,g_2)$ as follows: 
%%%%%%%%%%%%%%%%%%
\begin{eqnarray}
g_+&=&-g_1\cos2\phi-g_2\sin2\phi,\label{eq:g1}\\
g_\times&=&-g_1\sin2\phi+g_2\cos2\phi,\label{eq:g2}
\end{eqnarray}
%%%%%%%%%%%%%%%%%%
where $\phi$ is the polar angle. We obtain $g_+$ and $g_\times$ for
each radial bin by simply averaging these shears of individual source
galaxies without weighting. We estimate the mean and error using a
jackknife resampling technique. The shear profiles are computed in the
range of radii $\theta=[0\farcm8,\,20']$ with a bin size of
$\Delta\log\theta=0.2$.  Thus we have seven radial bins in total.

\begin{figure}
\epsscale{0.48}
\plotone{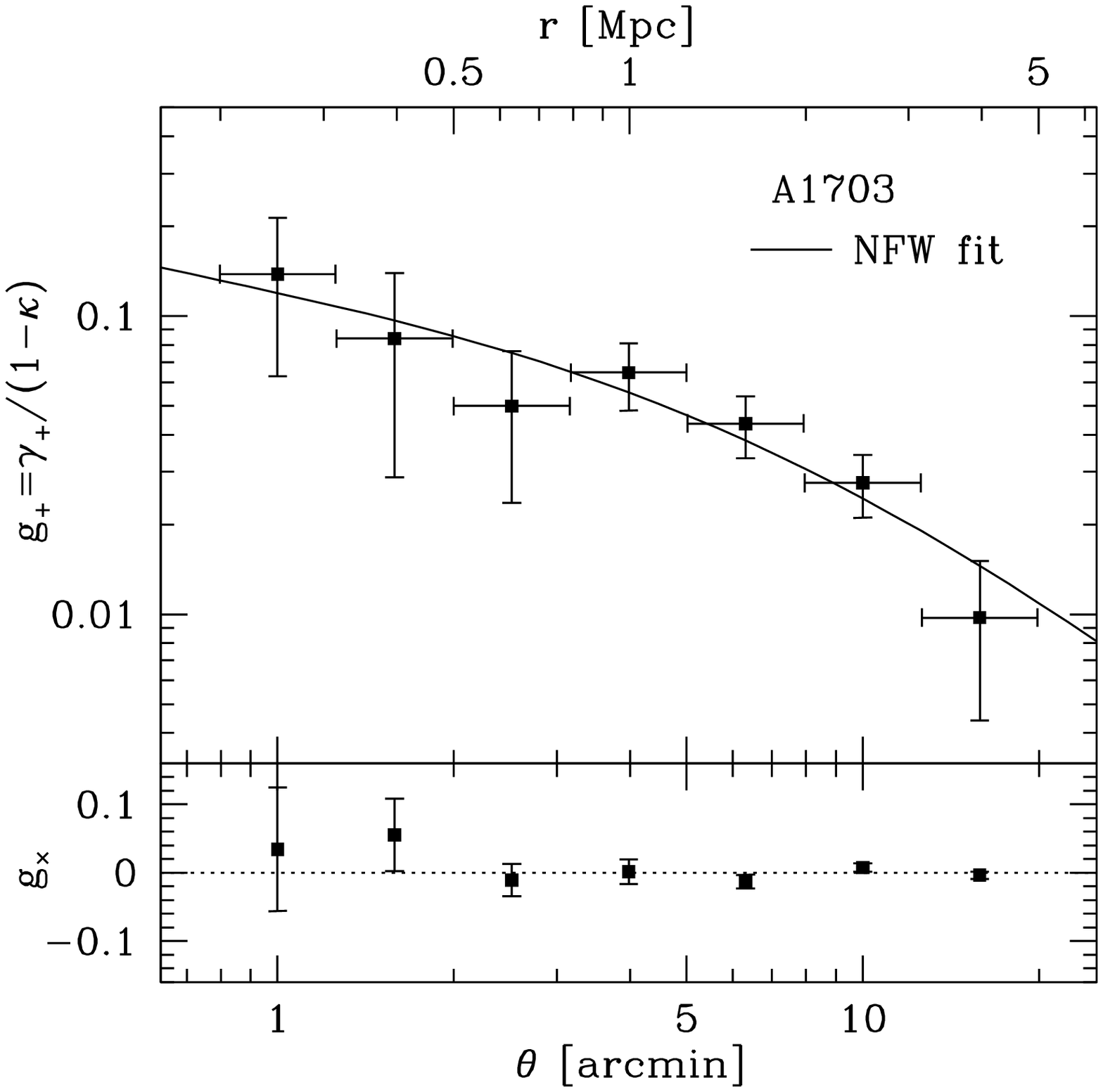}
\plotone{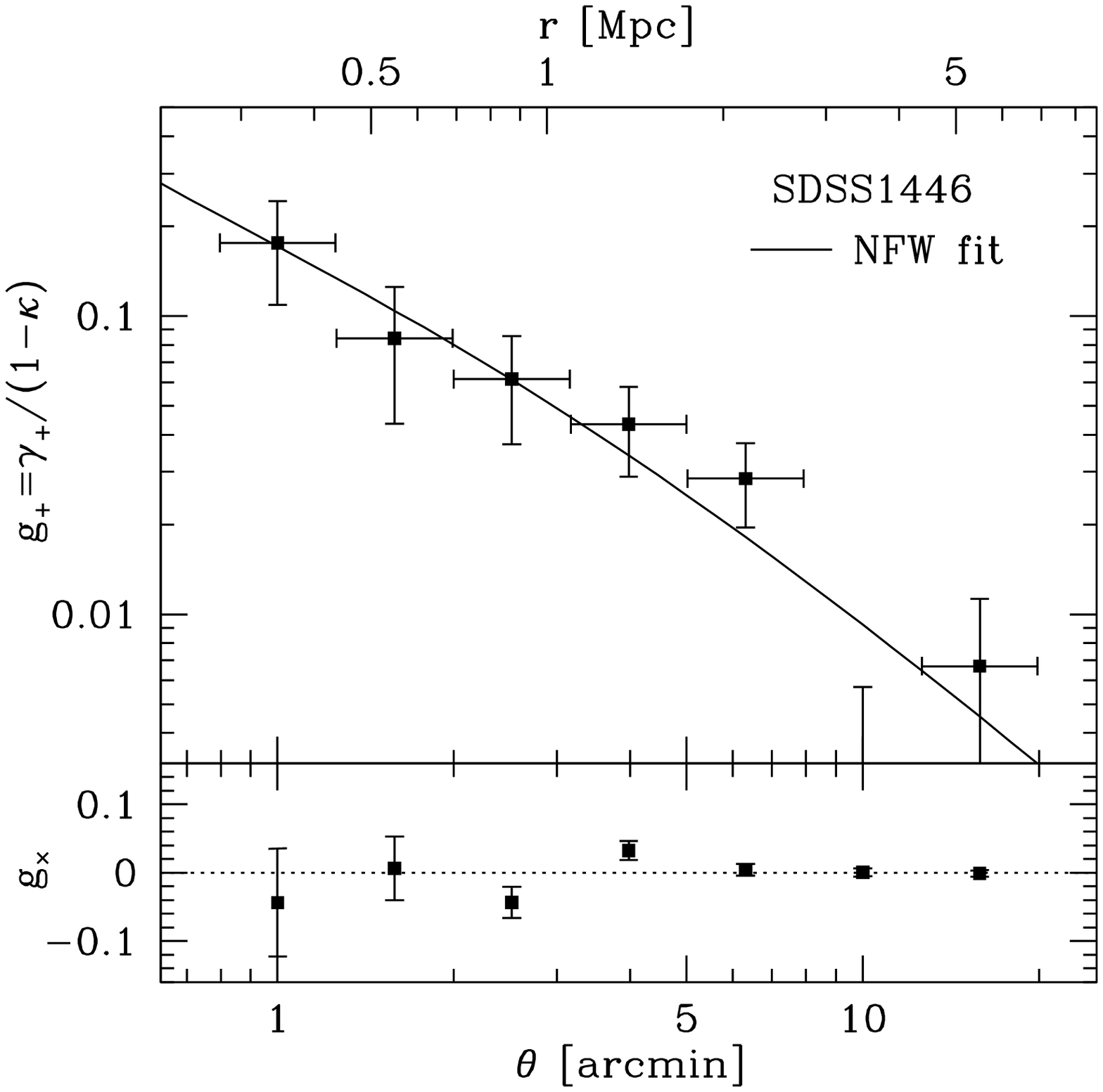} \\
\plotone{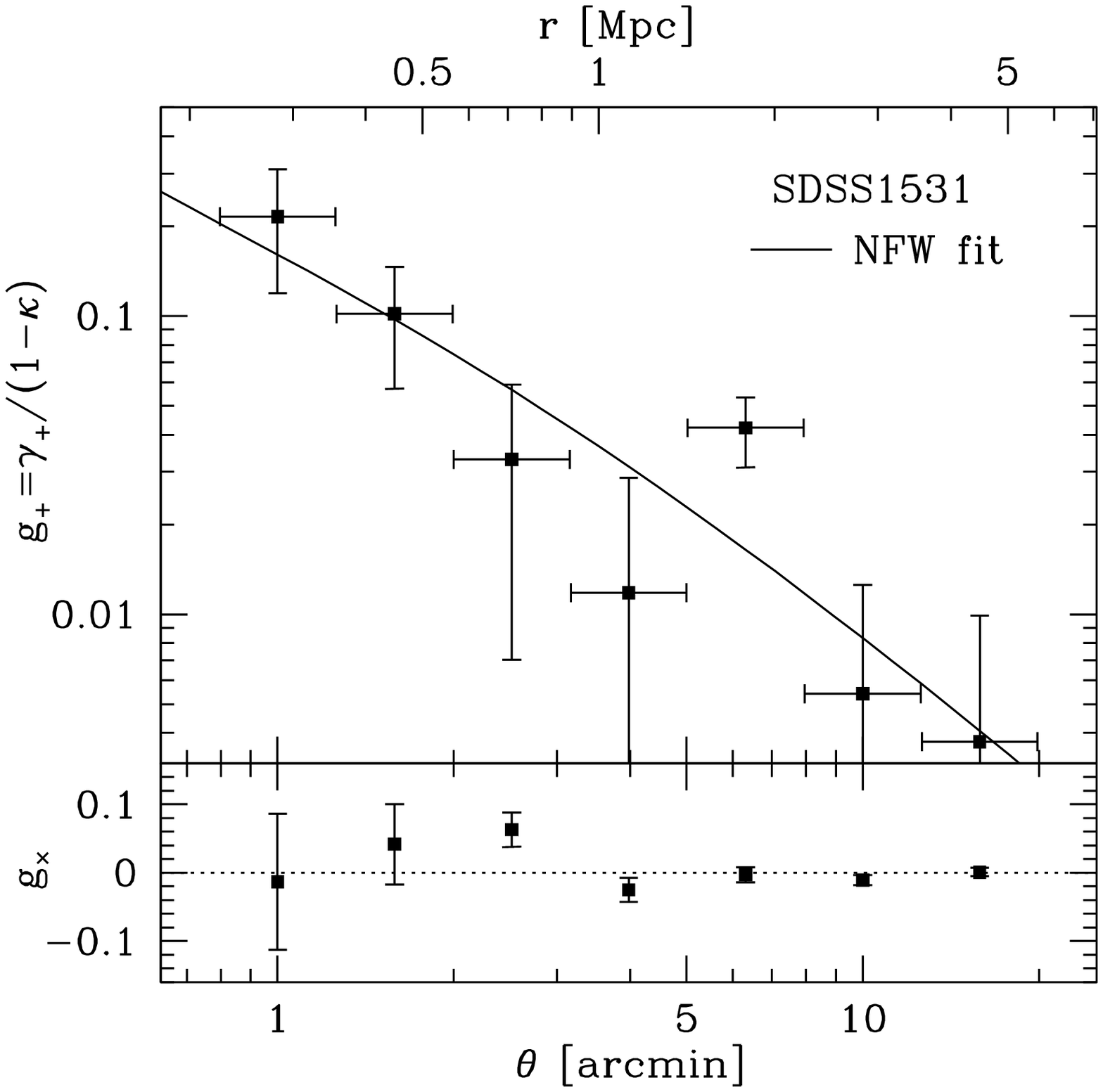}
\plotone{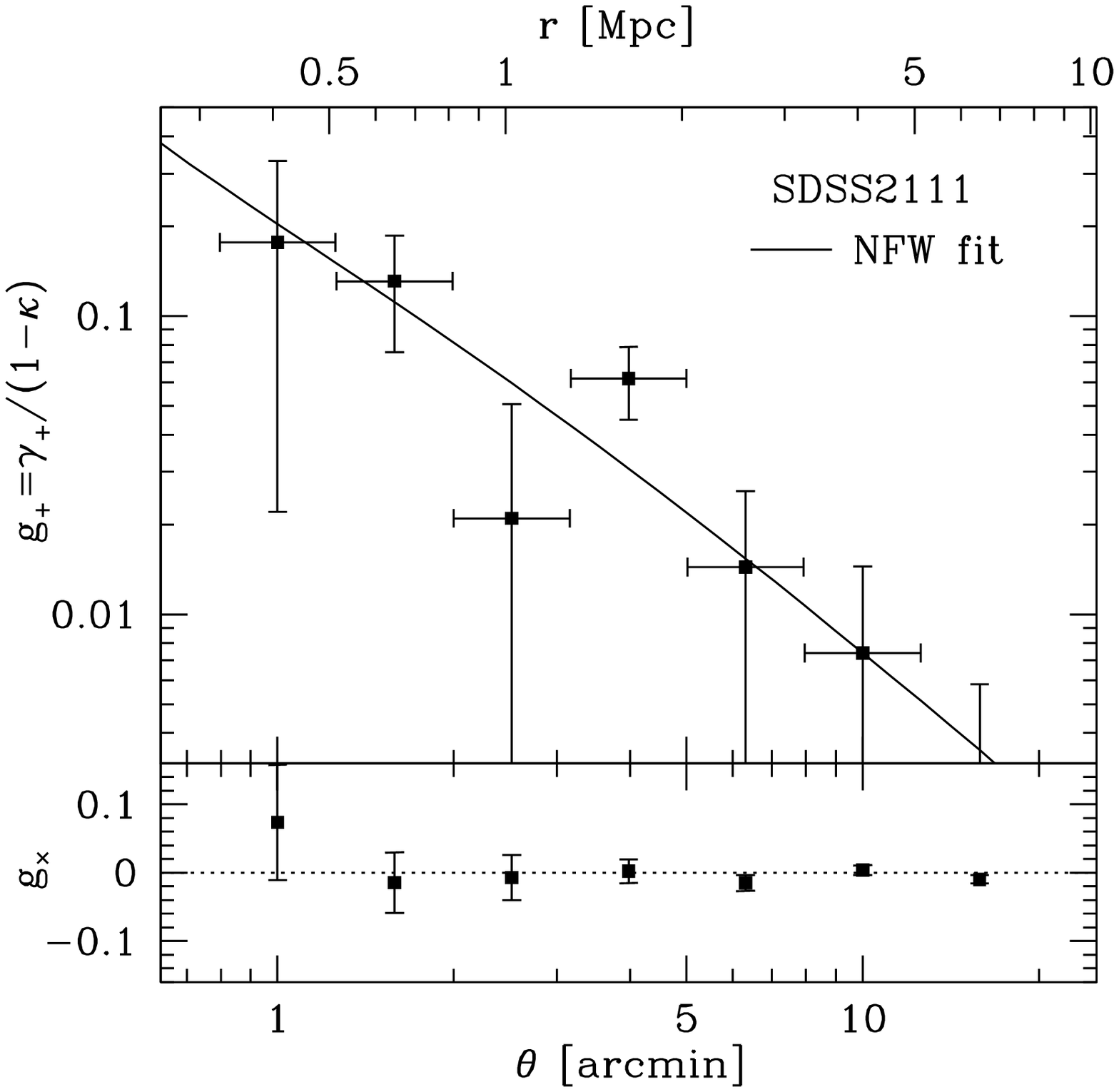}
\caption{Azimuthally-averaged tangential shear $g_+$
  (eq. [\ref{eq:g1}]) and the $45^\circ$ rotated component $g_\times$
  (eq. [\ref{eq:g2}]) as a function of distance from the cluster
  center. The NFW models fitted to the observed shear   profiles are
  shown by solid lines (see also Table~\ref{tab:swmodel}). 
\label{fig:profile_gamma}} 
\end{figure}

In Figure~\ref{fig:profile_gamma}, we plot radial profiles of $g_+$
and $g_\times$ of all the four clusters. We detect weak lensing
signals significantly up to $\sim 10'$ from the cluster center. On the
other hand, $g_\times$ is consistent with a null signal, as expected for
shears produced by gravitational lensing of clusters. 
We fit the tangential shear profile by the NFW profile
(eq. [\ref{eq:nfw}]) with the virial mass $M_{\rm vir}$ and
concentration parameters $c_{\rm vir}$ as free parameters.
For this, we adopt the following $\chi^2$:
%%%%%%%%%%%%%%%%%%
\begin{equation}
\chi^2_{\rm WL}=\sum_i\frac{\left[\bar{g}_{+,i}-g_+(\theta_i;\,M_{\rm
    vir},\,c_{\rm vir})\right]^2}{\sigma_i^2},
\label{eq:c2wl}
\end{equation}
%%%%%%%%%%%%%%%%%%
where $\bar{g}_{+,i}$ and $\sigma_i$ denote observed tangential shear and
its error for $i$-th radial bin. As shown in 
Figure~\ref{fig:profile_gamma}, the NFW model provides reasonable fits
to the data. The best-fit model parameters are summarized in
Table~\ref{tab:swmodel}.   

%%%%%%%%%%%%%%%%%%%%%%%%%%%%%%%%%%%%%%%%%%%%%%%%
%%%%%%%%%%%%%%%%%%%%%%%%%%%%%%%%%%%%%%%%%%%%%%%%
%%%%%%%%%%%%%%%%%%%%%%%%%%%%%%%%%%%%%%%%%%%%%%%%
\section{Combining Strong and Weak Lensing}
\label{sec:s+w}
%%%%%%%%%%%%%%%%%%%%%%%%%%%%%%%%%%%%%%%%%%%%%%%%
%%%%%%%%%%%%%%%%%%%%%%%%%%%%%%%%%%%%%%%%%%%%%%%%
%%%%%%%%%%%%%%%%%%%%%%%%%%%%%%%%%%%%%%%%%%%%%%%%

%%%%%%%%%%%%%%%%%%%%%%%%%%%%%%%%%%%%%%%%%%%%%%%%
\subsection{Constraints on Parameters}
%%%%%%%%%%%%%%%%%%%%%%%%%%%%%%%%%%%%%%%%%%%%%%%%

In this section, we combine results from strong lensing
(\S\ref{sec:slens}) and weak lensing (\S\ref{sec:wlens}) to constrain
the radial mass profiles over a wide range of radii. We do so simply
by summing up the chi-squares:
%%%%%%%%%%%%%%%%%%
\begin{equation}
\chi^2=\chi^2_{\rm SL}+\chi^2_{\rm WL},
\end{equation}
%%%%%%%%%%%%%%%%%%
where the weak lensing constraint $\chi^2_{\rm WL}$ is from equation 
(\ref{eq:c2wl}). Although we can in principle use strong lens
constraints obtained from detailed fitting of image positions in
\S\ref{sec:slens}, here we choose a conservative approach to adopt
only the Einstein radius as our strong lens constraint. Specifically,
$\chi^2_{\rm SL}$ is computed as
%%%%%%%%%%%%%%%%%%
\begin{equation}
\chi^2_{\rm SL}=\sum\frac{\left[\bar{\theta}_{\rm Ein}-\theta_{\rm Ein}(M_{\rm
    vir},\,c_{\rm vir})\right]^2}{\sigma_{\rm Ein}^2},
\end{equation}
%%%%%%%%%%%%%%%%%%
 where $\bar{\theta}_{\rm Ein}$ and $\sigma_{\rm Ein}$ are
 the Einstein radius and its errors listed in Table~\ref{tab:slein}.
 We include the asymmetry of errors on $\bar{\theta}_{\rm Ein}$ by
 using different values of $\sigma_{\rm Ein}$ between
 $\theta>\bar{\theta}_{\rm Ein}$ and $\theta<\bar{\theta}_{\rm Ein}$.
 For SDSS1446 and SDSS2111, only the allowed range of the redshift of
 the arcs (and hence the redshift of the Einstein radius, $z_{\rm
 Ein}$) was given; therefore, for these clusters we assume a flat
 prior $1.6<z_{\rm Ein}<3.5$ (SDSS1446) or $z_{\rm Ein}<3.5$
 (SDSS2111) and marginalize $\chi^2_{\rm SL}$ over $z_{\rm Ein}$ to
 derive strong lens constraints.  We note that the contribution of
 stars (member galaxies and the BCGs), which can be significant at
 strong lensing regime, is excluded in estimating $\theta_{\rm Ein}$
 from strong lens modeling. Thus the derived radial profile should be
 considered as that for the sum of dark matter and intracluster gas
 rather than the total matter profile. 

\begin{figure}
\epsscale{0.48}
\plotone{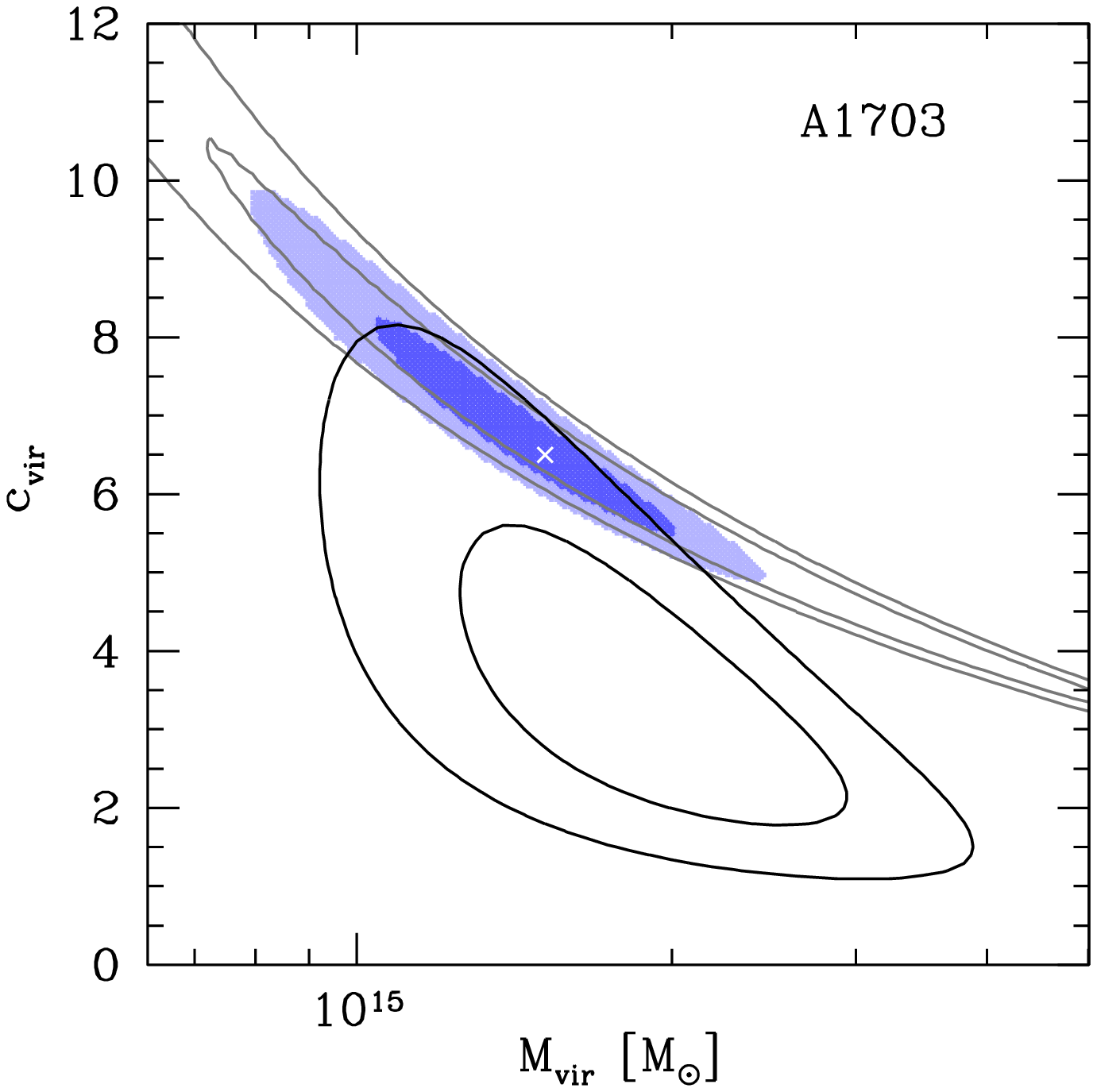}
\plotone{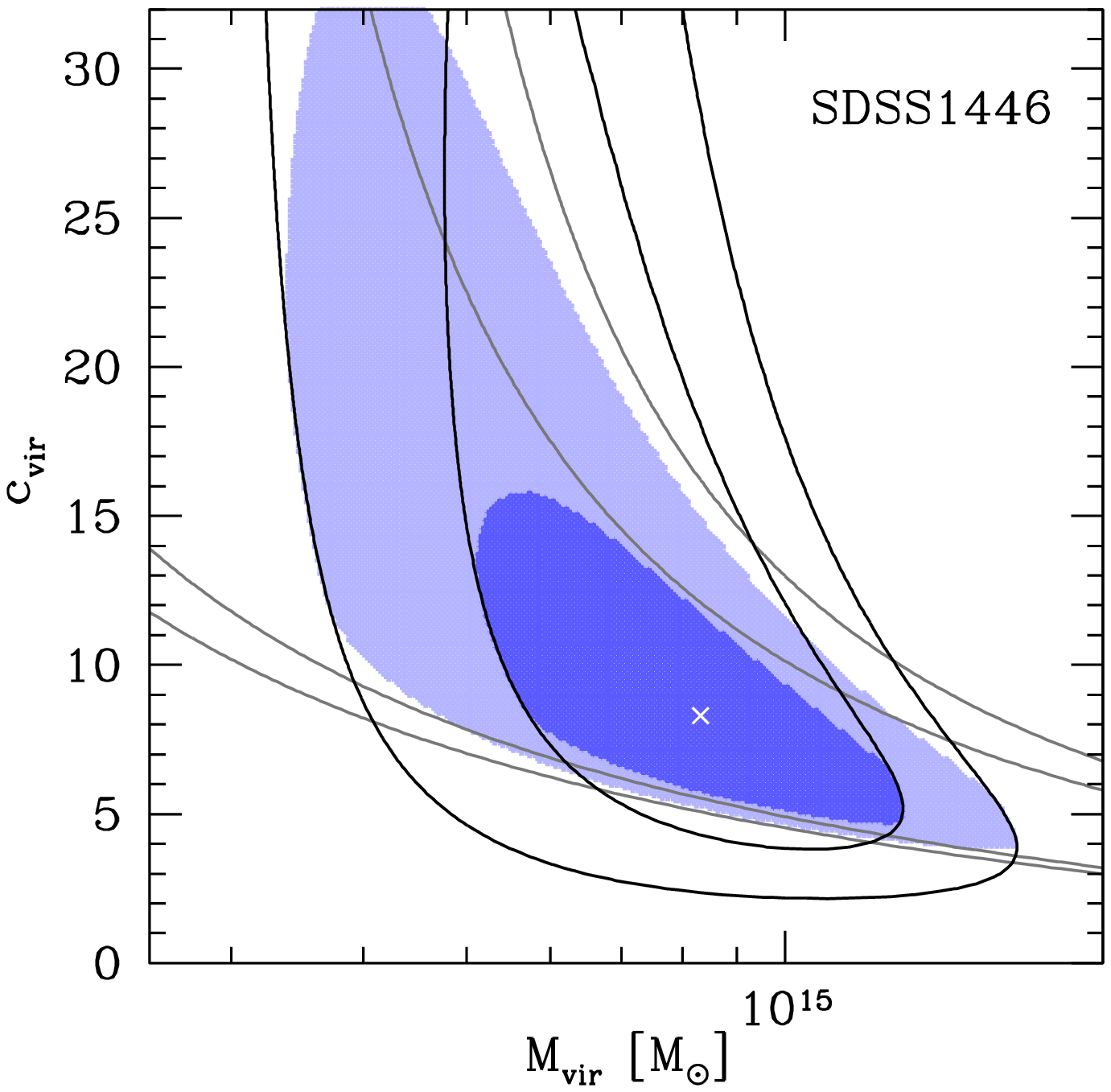} \\
\plotone{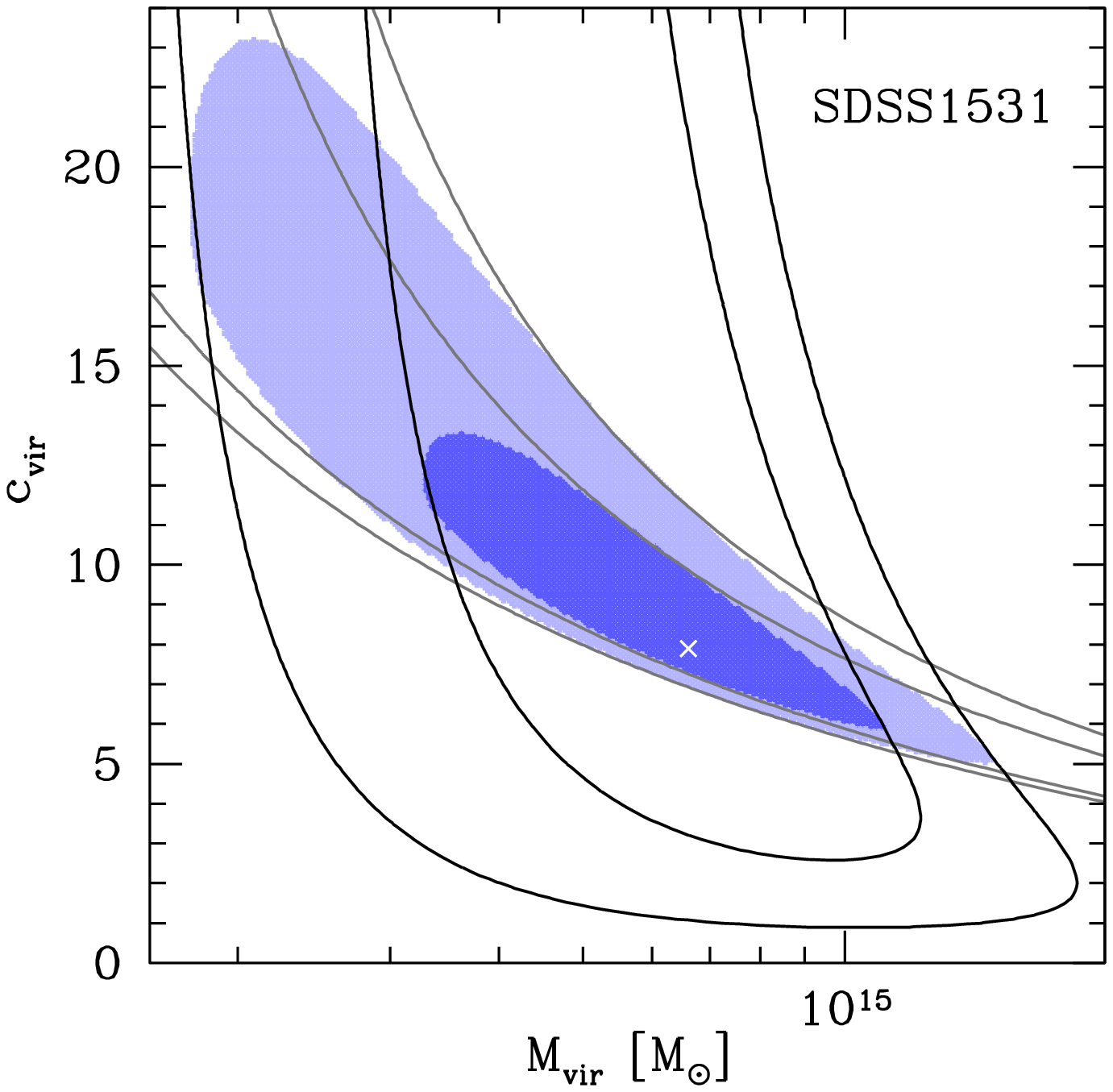}
\plotone{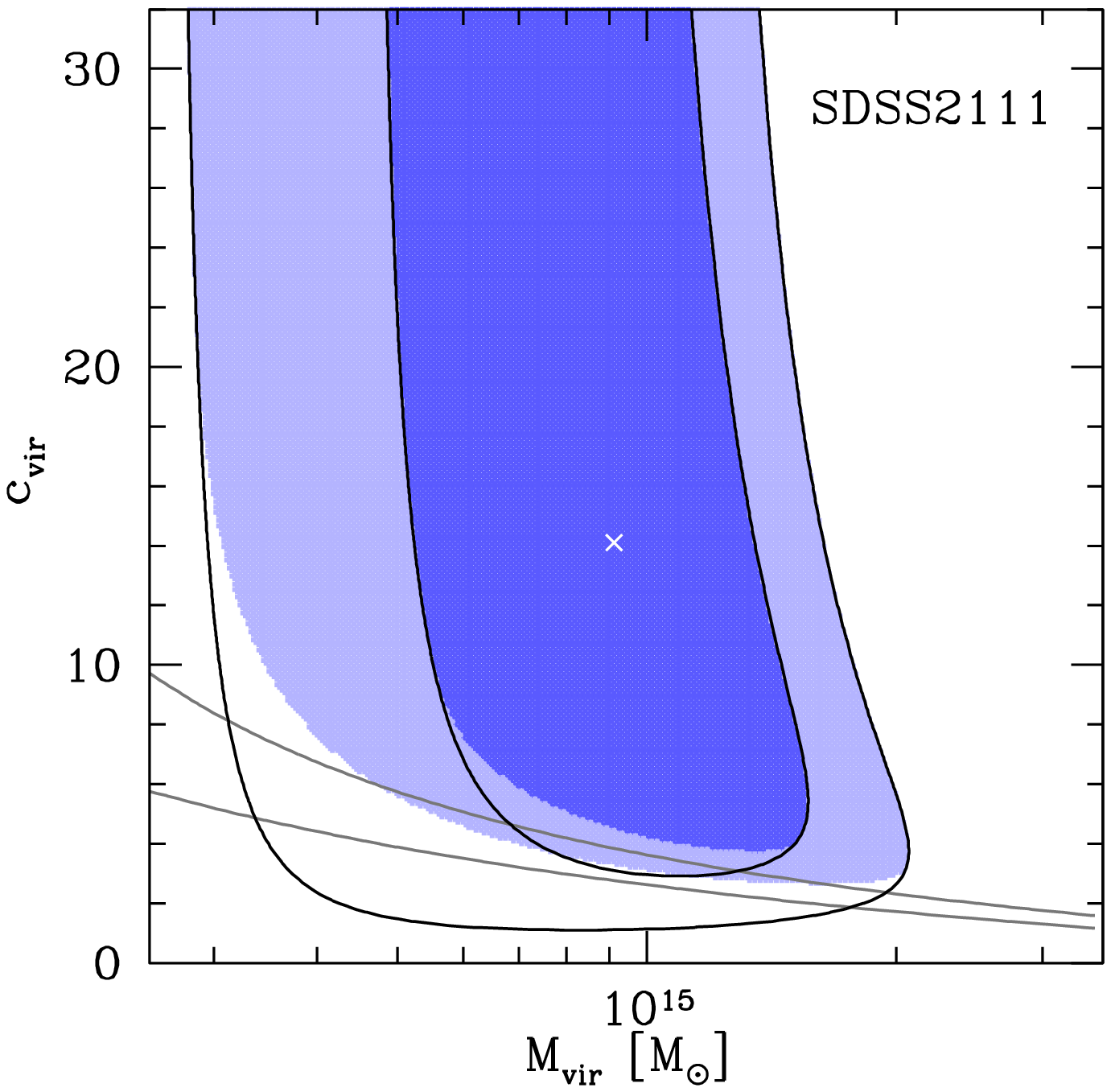}
\caption{Contours at 1$\sigma$ and 2$\sigma$ confidence levels in the
  $M_{\rm vir}$-$c_{\rm vir}$ plane. Black and grey lines indicate
  constraints from  weak or strong lensing, respectively. Combined
  strong and weak lensing constraints are plotted by shaded regions. 
  The best-fit model parameters from combined strong and weak lensing
  are shown by crosses (see also Table~\ref{tab:swmodel}).  
\label{fig:cont}} 
\end{figure}

We show the constraints in the $M_{\rm vir}$-$c_{\rm vir}$ plane in
Figure~\ref{fig:cont}. As expected, adding strong lensing
significantly improve constraints on these parameters, especially the
concentration parameter $c_{\rm vir}$, although this is not the case
for SDSS2111 because of the unknown arc redshift and the relatively
large error on $\theta_{\rm Ein}$. Constraints from strong and
weak lensing are consistent with each other, except A1703 for which
the best-fit models slightly differ. We list the best-fit model
parameters in Table~\ref{tab:swmodel}. It is found that the radial
profiles are fitted well by the NFW profile with $\chi^2/{\rm dof}\sim 1$.

\begin{figure}
\epsscale{0.48}
\plotone{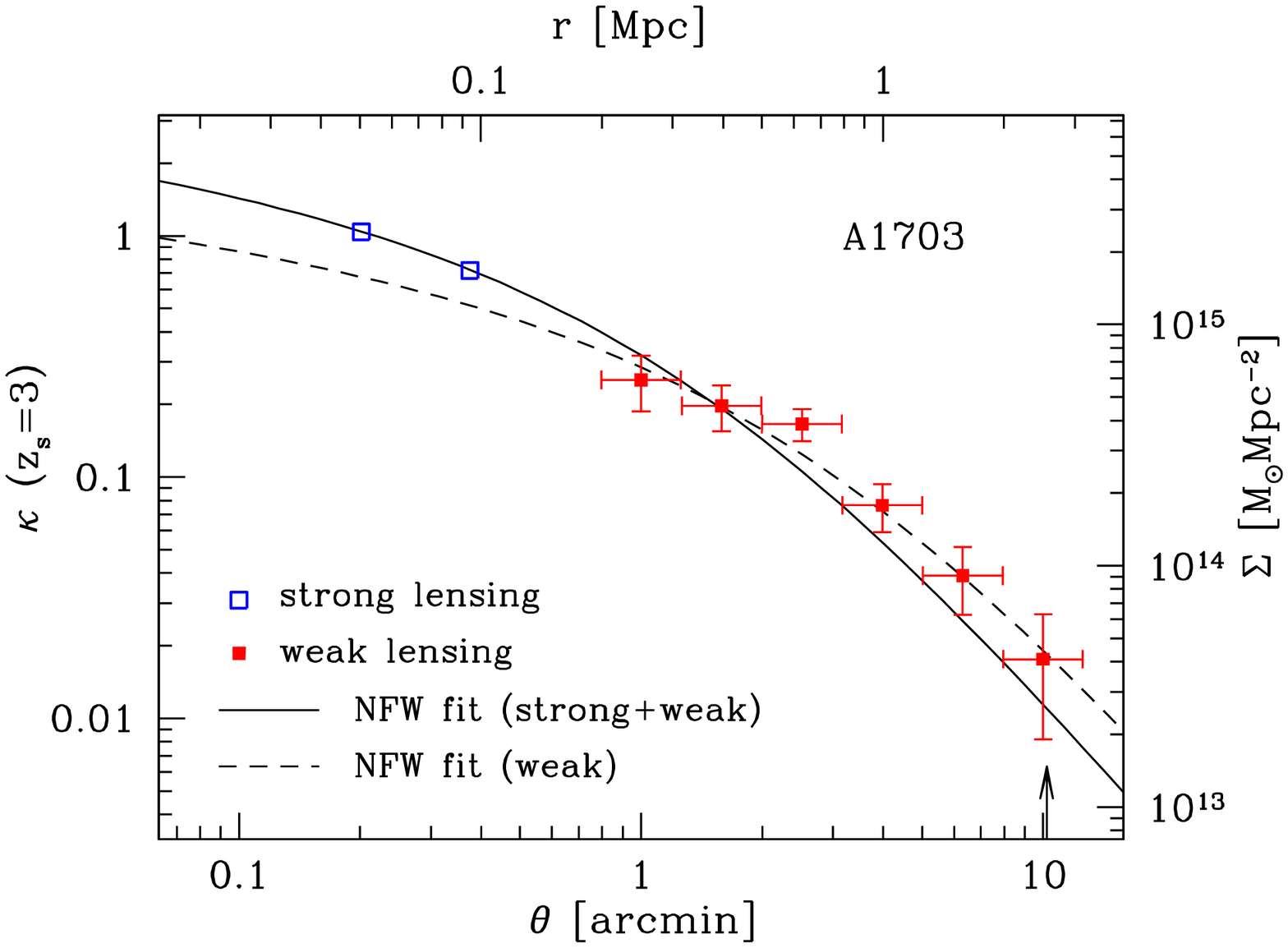}
\plotone{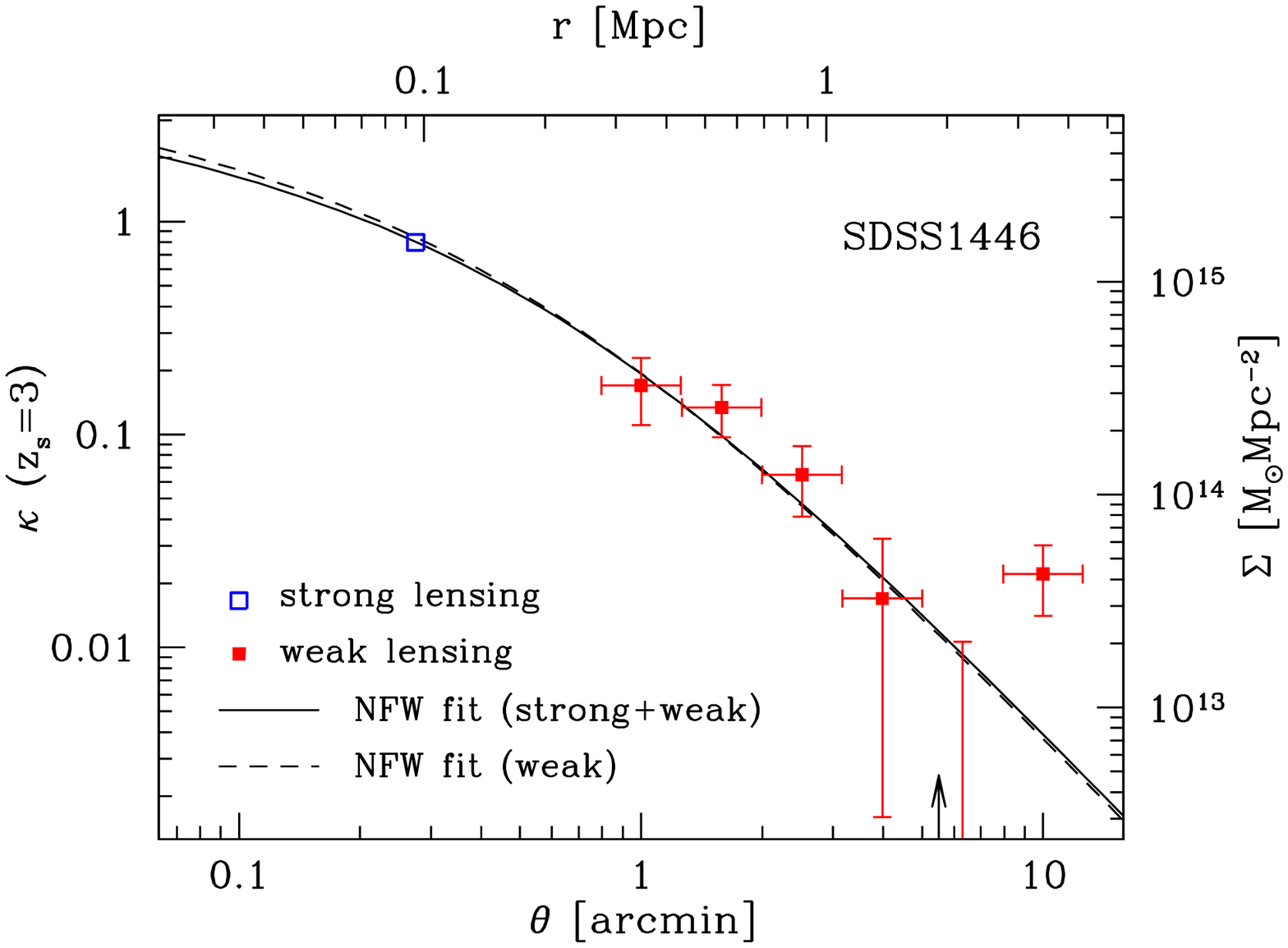} \\
\plotone{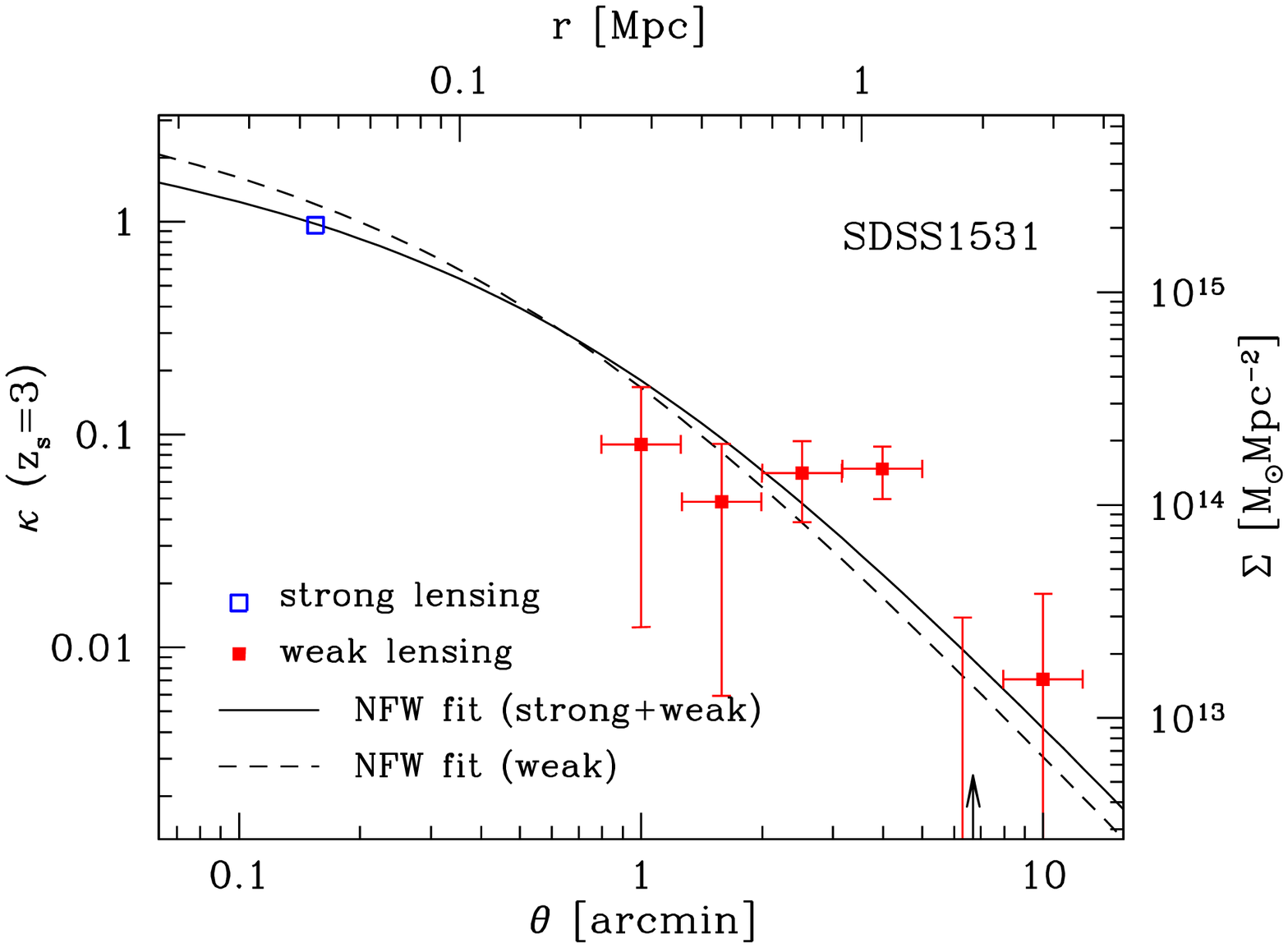}
\plotone{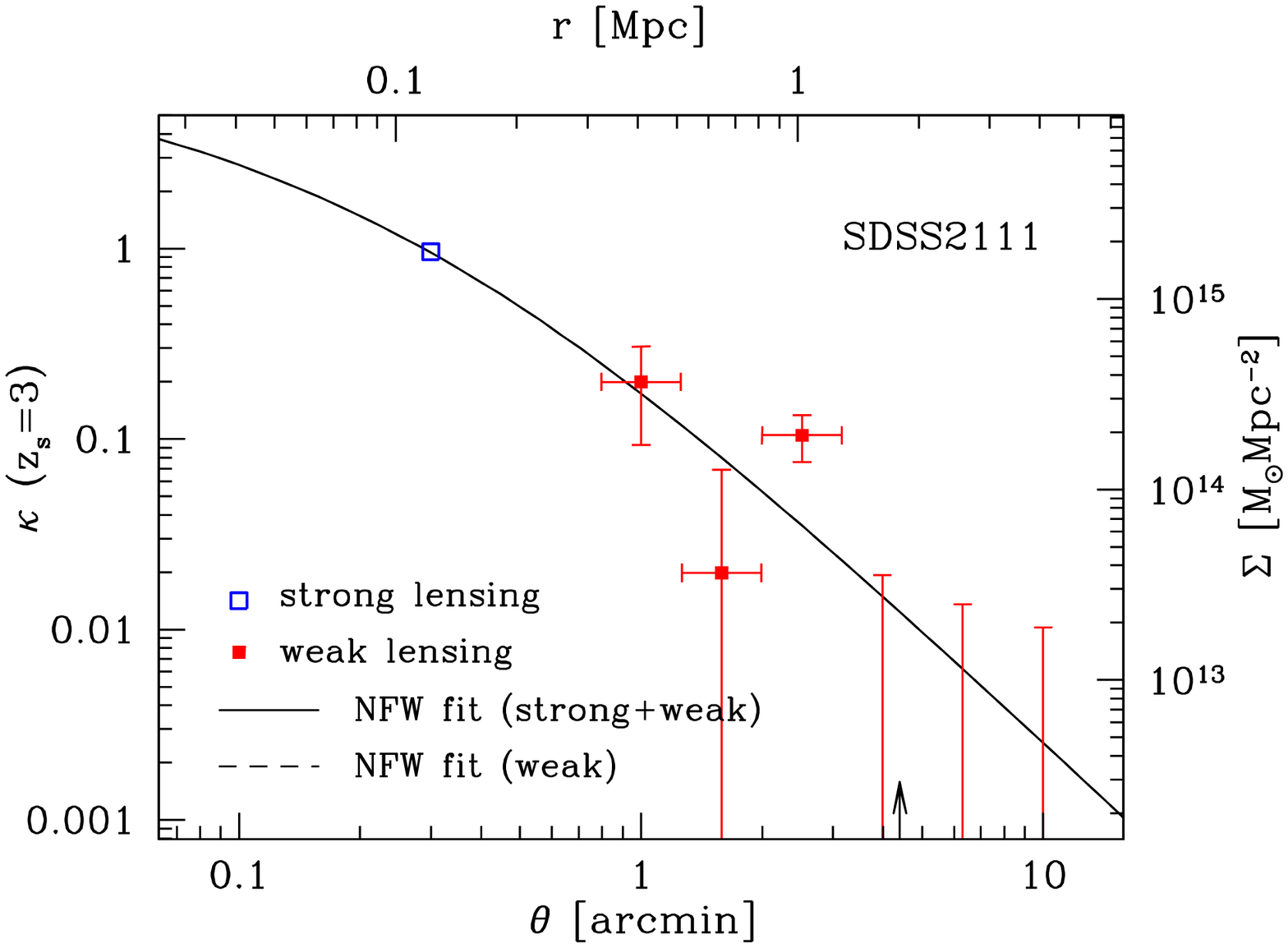}
\caption{Radial profiles of convergence $\kappa$ for the source
  redshift $z_s=3$ constrained from strong and weak lensing. Radial
  profiles of best-fit models from combined strong and weak lensing
  are plotted by solid lines, whereas those from weak lensing only are
  shown by dashed lines  (see also Table~\ref{tab:swmodel}).  Filled
  squares are $\kappa$ profiles reconstructed from weak lensing
  tangential shear profiles shown in Figure~\ref{fig:profile_gamma};
  as boundary conditions, we assumed values of convergences in the
  outermost radial bin to be those computed from the best-fit NFW
  profiles from weak lensing. We indicate the Einstein radii used as
  strong lens constraints by open squares, and the virial radii of the
  best-fit models from strong and weak lensing analysis by arrows.
  \label{fig:profile_kappa}} 
\end{figure}

To see how strong and weak lensing probe different radii, we plot
radial profiles of convergence $\kappa$ from lensing observations and
best-fit models in Figure~\ref{fig:profile_kappa}. We derive
convergence at each radial bin from weak lensing shears by using the
following relation: 
%%%%%%%%%%%%%%%%%%
\begin{equation}
g_+(\theta)\left[1-\kappa(\theta)\right]=\frac{2}{\theta^2}
\int_0^\theta\theta'\kappa(\theta')d\theta'-\kappa(\theta).
\end{equation}
%%%%%%%%%%%%%%%%%%
We need a boundary condition in order to solve this equation. We fix
$\kappa$ in the outermost bin to the value computed from the NFW
profile with best-fit parameter values from weak lensing data
alone. See, e.g., \citet{umetsu08} for practical procedures to compute
$\kappa$ from discrete $g_+$ data. As clearly shown in the Figure, 
strong lensing probes radii roughly an order of magnitude smaller than 
those constrained by weak lensing. Hence by combining strong and weak
lensing we can constrain the density profile of the cluster over 2~dex
in radius. We find that the model that fits weak lensing data of A1703
underpredicts convergence in the core of the cluster; the Einstein
radius of A1703 implies more centrally concentrated profile of A1703
than expected from weak lensing data alone. For the other three
clusters, convergence profiles from weak lensing alone and strong and
weak lensing agree quite well. However, strong lensing data do narrow
down the allowed range of radial profiles, as is clear from
Figure~\ref{fig:cont}.  

We now compare our result of A1703 with the combined strong and weak
lensing analysis by \citet{broadhurst08a}. We find that the best-fit
virial masses are consistent with each other, but the concentration
parameter of our best-fit model, $c_{\rm vir}=6.5_{-0.7}^{+1.2}$,
differs from their best-fit value, $c_{\rm vir}=9.9_{-1.6}^{+2.4}$ at 
$\sim 2\sigma$ level. We ascribe the difference to the larger Einstein
radius they assumed, $\theta_{\rm Ein}=33''$ for $z_s=2.8$. Indeed,
our best-fit model predicts $\theta_{\rm Ein}=22''$ for $z_s=2.8$,
which appears more consistent with the locations of lensed arcs in
A1703 \citep[see also][]{richard09}. 

%%%%%%%%%%%%%%%%%%%%%%%%%%%%%%%%%%%%%%%%%%%%%%%%
\subsection{Distribution of the Concentration Parameter}
%%%%%%%%%%%%%%%%%%%%%%%%%%%%%%%%%%%%%%%%%%%%%%%%

One of our main interests lies in the possible excess of the
concentration parameter $c_{\rm vir}$ found among lensing
clusters. Here we compare our results with theoretical predictions
based on the $\Lambda$CDM model.

For the theoretical prediction, we adopt results of $N$-body
simulations in the Wilkinson Microwave Anisotropy Probe (WMAP) 5-year
cosmology \citep{duffy08}. They derived the mean concentration of dark
halos as a function of halo mass and redshift as 
%%%%%%%%%%%%%%%%%%
\begin{equation}
\bar{c}_{\rm vir}({\rm sim})=\frac{7.85}{(1+z)^{0.71}}\left(\frac{M_{\rm
      vir}}{2.78\times 10^{12}M_\odot}\right)^{-0.081}.
\label{eq:c_duffy}
\end{equation}
%%%%%%%%%%%%%%%%%%
The apparent concentration parameters for halos with given mass and
redshift are significantly scattered, with approximately the log-normal
distribution with $\sigma_{\log c}\simeq 0.14$. However, the diversity
of the halo population found in $N$-body simulations suggest that
clusters with giant arcs should represent a quite biased population. 
This lensing bias arises from the fact that the lensing cross section
is a sensitive function of the halo concentration. Therefore we expect 
projected mass distributions of lensing clusters to be more centrally
concentrated than ordinary clusters, implying that the clusters are
intrinsically more concentrated and/or their major axes are
preferentially aligned with the line-of-sight, giving rise to
apparently more concentrated distributions of projected surface mass
densities.  \citet{hennawi07} estimated the lensing clusters have
$\sim 40\%$ higher values of concentrations compared with normal
clusters \citep[see also][]{fedeli07}. \citet{oguri09} focused on
clusters with larger Einstein radii and argued that such clusters have
$\sim 40-60\%$ higher concentrations. Based on these discussions, in
this paper we consider $50\%$ enhancement of $c_{\rm vir}$ due to the
lensing bias. Note that the lensing bias includes the projection
effect from the halo triaxiality. From the calculation of \citet{oguri09}, it
is also found that the apparent concentration of the lensing clusters
has slightly smaller scatter with $\sigma_{\log c}\simeq 0.12$.

\begin{figure}
\epsscale{0.6}
\plotone{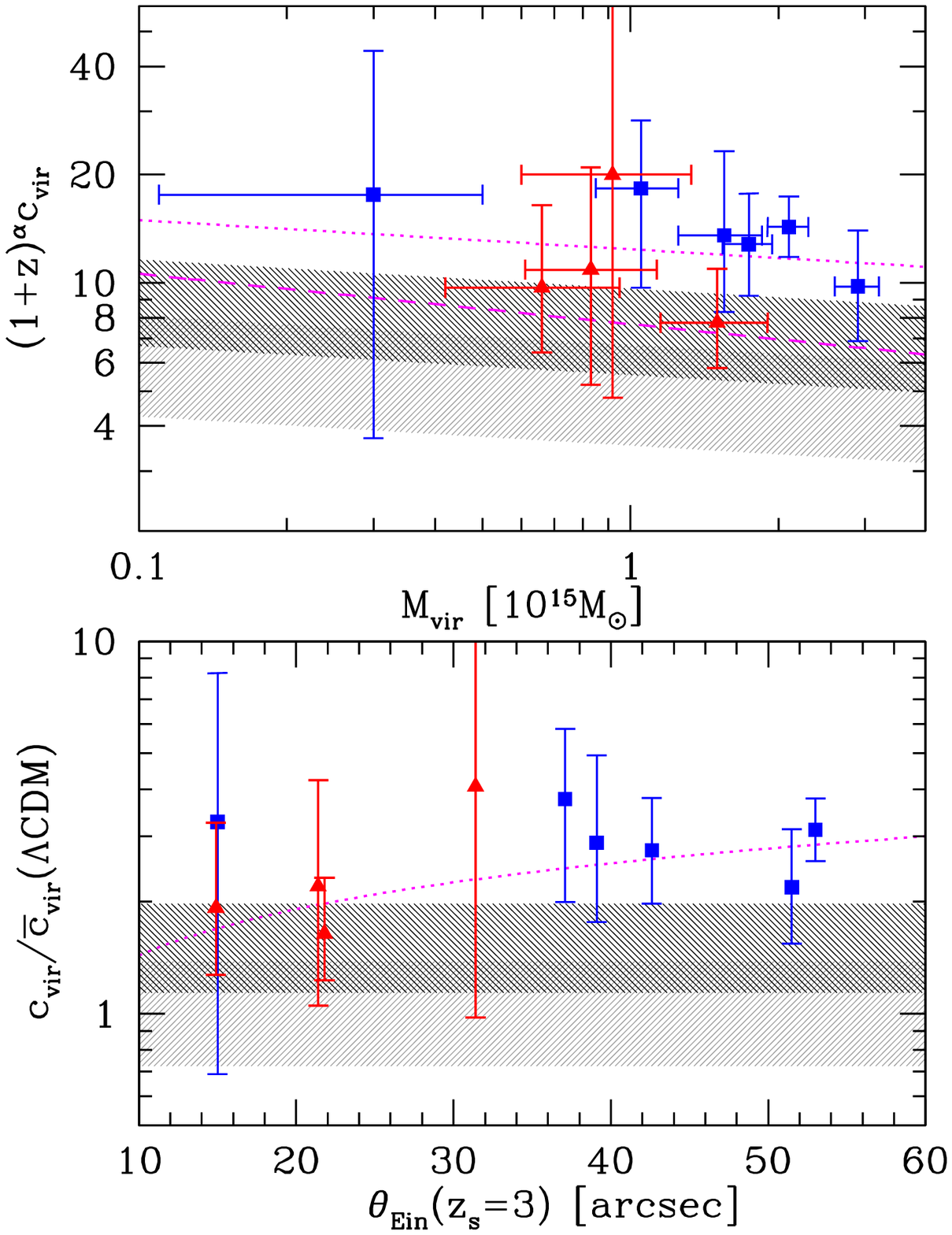}
\caption{Distribution of the concentration parameter $c_{\rm vir}$
  constrained from combined strong and weak lensing analysis. Filled
  triangles show our results presented in this paper, whereas filled
  squares are from literature, results for A1689 \citep{umetsu08},
  A2261 \citep{umetsu09}, A370, RXJ1347,
  CL0024 \citep{broadhurst08a}, and MS2137 \citep{gavazzi03}. 
  {\it Upper:} The distribution of $c_{\rm vir}$ as a function of
  virial mass $M_{\rm vir}$. The data points are corrected for the
  redshift evolution of the concentration parameter predicted in the
  $\Lambda$CDM model, $c_{\rm vir}\propto (1+z)^{-\alpha}$ with $\alpha=0.71$
  \citep{duffy08}. The grey shaded regions indicate 1$\sigma$ range of
  $c_{\rm vir}$ derived from $\Lambda$CDM simulations by \citet{duffy08}. 
  Black shaded regions show predicted concentration parameters after
  approximately taking the lensing bias ($\sim 50\%$) into account
  \citep{hennawi07,oguri09}. The dashed line plots the $M_{\rm
  vir}$-$c_{\rm vir}$ relation determined observationally from a
  sample of lensing and X-ray clusters \citep{comerford07}. The dotted
  line shows the best-fit curve to the data.
  {\it Lower:} The ratios of $c_{\rm vir}$
  from analyses of lensing clusters to those expected in $\Lambda$CDM 
  are plotted a function of the Einstein radius $\theta_{\rm Ein}$ for
  the source redshifts $z_s=3$. Again, the shaded regions show
  1$\sigma$ range expected in $\Lambda$CDM, with ({\it black}) and
  without ({\it grey}) the lensing bias. The best-fit curve by a
  power-law is shown by the dotted line.
  \label{fig:clupar}} 
\end{figure}

Figure~\ref{fig:clupar} shows the distribution of $c_{\rm vir}$ of
the four clusters we study in this paper as well as several lensing
clusters studied before. We consider only clusters whose radial
profiles are well constrained from combined strong and weak lensing
analyses. We find that all the four clusters have best-fit value of
$c_{\rm vir}$ higher than the theoretical expectations. We still see the
excess even if we take the lensing bias into account, although the
value of $c_{\rm vir}$ for each cluster is marginally consistent with
the theory if the error is taken into account. Put another way, the
excess is not so strong as claimed in earlier work based on analysis
of different lensing clusters
\citep{broadhurst05a,broadhurst08a,comerford07}. By combining all the
10 clusters, it is clear that our sample of lensing clusters has
larger concentrations than expected from the $M_{\rm vir}$-$c_{\rm vir}$ 
relation predicted by the $\Lambda$CDM model. The concentrations are
also higher than the relation determined observationally by
\citet{comerford07} for a sample of lensing and X-ray clusters. Since
our sample cannot constrain the redshift and mass dependence of the
mean concentration very well, we fix them to those in equation
(\ref{eq:c_duffy}) and fit the overall normalization to the data. We
find   
%%%%%%%%%%%%%%%%%%
\begin{equation}
\bar{c}_{\rm vir}({\rm fit})=\frac{12.4}{(1+z)^{0.71}}\left(\frac{M_{\rm
      vir}}{10^{15}M_\odot}\right)^{-0.081},
\end{equation}
%%%%%%%%%%%%%%%%%%
as our best-fit to the results of 10 lensing clusters. The data are
inconsistent with the concentration parameter predicted in the
$\Lambda$CDM by $7\sigma$, even if we include the $50\%$ enhancement
to account for the lensing bias.\footnote{The conclusion is unchanged
  even if we relax the lower limit of the arc redshift of SDSS1446
  which originated from the absense of [OII] emission line. This is
  because the lower limit of the arc redshift corresponds to the upper  
  limit of the cluster core mass and hence to the upper limit of the
  concentration parameter. }   A simple average of theconcentration
parameter (with the inverse of the measurement error as a weight)
ignoring the mass and redshift dependence is $\langle c_{\rm
  vir}\rangle =9.3\pm2.6$.   

We also check the excess as a function of the Einstein radius, which
is a central quantity to characterize the central structure of a
cluster \citep[e.g.,][]{broadhurst08b}.  The plot in
Figure~\ref{fig:clupar} shows that the excess of $c_{\rm vir}$
depends slightly on the Einstein radius. We fit the data by a
power-law and find 
%%%%%%%%%%%%%%%%%%
\begin{equation}
\frac{\bar{c}_{\rm vir}({\rm fit})}{\bar{c}_{\rm vir}({\rm
    sim})}=2.4\left(\frac{\theta_{\rm Ein}}{35''}\right)^{0.41},
\end{equation}
%%%%%%%%%%%%%%%%%%
where $\theta_{\rm Ein}$ is the Einstein radius for the source
redshift $z_s=3$. The hypothesis that this ratio does not depend on
the Einstein radius is rejected at 99\% confidence level.
The weak dependence of the excess on the Einstein radius is
reasonable in the sense that concentrated two-dimensional mass
distributions are required for clusters to produce large Einstein
radii \citep[e.g.,][]{broadhurst08a,oguri09}.  

%%%%%%%%%%%%%%%%%%%%%%%%%%%%%%%%%%%%%%%%%%%%%%%%
%%%%%%%%%%%%%%%%%%%%%%%%%%%%%%%%%%%%%%%%%%%%%%%%
%%%%%%%%%%%%%%%%%%%%%%%%%%%%%%%%%%%%%%%%%%%%%%%%
\section{Summary and Discussion}
\label{sec:summary}
%%%%%%%%%%%%%%%%%%%%%%%%%%%%%%%%%%%%%%%%%%%%%%%%
%%%%%%%%%%%%%%%%%%%%%%%%%%%%%%%%%%%%%%%%%%%%%%%%
%%%%%%%%%%%%%%%%%%%%%%%%%%%%%%%%%%%%%%%%%%%%%%%%

In this paper, we have studied the mass profiles of four clusters
by combining strong lens constraints with weak lensing shear
measurements. We have drawn our sample from giant arc clusters newly
discovered by the SGAS, the new arc survey using the SDSS data
\citep{hennawi08}.  We take advantage of follow-up wide-field {\it
  Subaru} Suprime-cam images to obtain weak lensing constraints out 
to nearly the virial radii of the clusters. The central densities of
the clusters are determined well from the arcs with the redshifts
spectroscopically measured partly from our ongoing program with the
{\it Gemini} telescope. The technique allows us to study the radial
mass profile over 2~dex in radius, which is essential for reliable
extractions of concentrations of the clusters. 

We have found that the radial profiles are fitted well by the NFW
profile. We have determined the virial mass $M_{\rm vir}$ and
concentration parameter $c_{\rm vir}$ of the four clusters accurately
from the combined strong and weak lensing analysis. We confirmed that
strongly lensed background galaxies with measured redshifts indeed
improve constraints on mass profiles, particularly $c_{\rm vir}$. We have
found the values of $c_{\rm vir}$ for our 4 clusters to be
$c_{\rm vir}\sim 8$, except for SDSS2111 whose radial profile was not
constrained very well because of insufficient strong lens
information. The values are slightly higher than the $\Lambda$CDM
predictions, even if we take account of the lensing bias that clusters 
with giant arcs are more centrally concentrated (intrinsically and/or
apparently due to the projection effect) than normal clusters,
although the excess is not as large as that claimed in earlier
work. By combining all the 10 clusters with strong plus weak lensing
analysis available, we confirm a $7\sigma$ excess of the concentration
parameter compared with the $\Lambda$CDM prediction. We find that the
excess is dependent on the Einstein radius of the system such that
clusters with larger Einstein radii show larger excess of the
concentration parameters.   

There are several possible explanations for the excess of the
concentration parameter. One such explanation is the effect of
baryons
\citep[e.g.,][]{kazantzidis04,gnedin04,puchwein05,lin06,rozo08}. 
In particular,
the adiabatic contraction associated with the baryon cooling can
enhance the core density of dark matter and hence can increase the
concentration of clusters. Another possibility is that theoretical
predictions are not so accurate. In particular, the probability
distribution of the concentration parameter for very massive halos
($\gtrsim 10^{15}h^{-1}M_\odot$) and its redshift evolution has not
been studied very much in $N$-body simulations, and thus our
theoretical predictions inevitably rely on extrapolations from
lower-mass halos. For example, it has been claimed that the redshift
evolution of the concentrations of most massive halos may be
different from that of less massive halos \citep[e.g.,][]{zhao03}. 
Thus it is important to perform many realizations of large box-size
$N$-body simulations to improve the accuracy of theoretical
predictions at the high mass end. A more exotic interpretation is that
clusters form earlier than expected from the $\Lambda$CDM model,
as the concentration is known to correlate with the formation epoch of
the cluster \citep{wechsler02}. Such modification of the formation
epoch can for instance be realized by considering early dark energy or
primordial non-Gaussianity
\citep[e.g.,][]{mathis04,sadeh07,sadeh08,oguri09}.  

This paper has presented initial results of detailed studies for our
unique sample of giant arc clusters \citep{hennawi08}. To investigate
the structure of lensing clusters in a more systematic and statistical
manner, it is of great importance to extend this research by applying
the technique we have developed in this paper to other lensing
clusters.  In addition, it is important to improve constraints on the
mass models of individual clusters by adding  more data. Measuring
redshifts of more arcs will significantly refine our strong lens
modeling. In addition, dynamical information from the velocity dispersion
measurement, as well as X-ray and Sunyaev-Zel'dovich signals, provide
an important cross check of our mass models
\citep[e.g.,][]{mahdavi07,lemze08,lemze09}. We are planning these
follow-up observations for our sample of giant arc clusters.  

\acknowledgments

We thank Marceau Limousin and Johan Richard for useful correspondence
regarding the cluster A1703, Tom Broadhurst, Keiichi Umetsu, Elinor
Medezinski, Masahiro Takada, Nobuhiro Okabe, and Maru{\v s}a 
Brada{\v c} for discussions on weak lensing, and Gilles Orban de
Xivry, Phil Marshall, and Roger Blandford for for bringing our
attention to a higher-order catastrophe. 
This work was supported in part by Department of Energy contract
DE-AC02-76SF00515. The authors wish to recognize and acknowledge the
very significant cultural role and reverence that the summit of Mauna
Kea has always had within the indigenous Hawaiian community.  We are
most fortunate to have the opportunity to conduct observations from
this mountain.

\clearpage

\appendix

%%%%%%%%%%%%%%%%%%%%%%%%%%%%%%%%%%%%%%%%%%%%%%%%
%%%%%%%%%%%%%%%%%%%%%%%%%%%%%%%%%%%%%%%%%%%%%%%%
%%%%%%%%%%%%%%%%%%%%%%%%%%%%%%%%%%%%%%%%%%%%%%%%
\section{Strong Lensing Modeling of A1703 by a Generalized 
NFW Profile}
\label{sec:gnfw}
%%%%%%%%%%%%%%%%%%%%%%%%%%%%%%%%%%%%%%%%%%%%%%%%
%%%%%%%%%%%%%%%%%%%%%%%%%%%%%%%%%%%%%%%%%%%%%%%%
%%%%%%%%%%%%%%%%%%%%%%%%%%%%%%%%%%%%%%%%%%%%%%%%

\begin{figure}
\epsscale{0.5}
\plotone{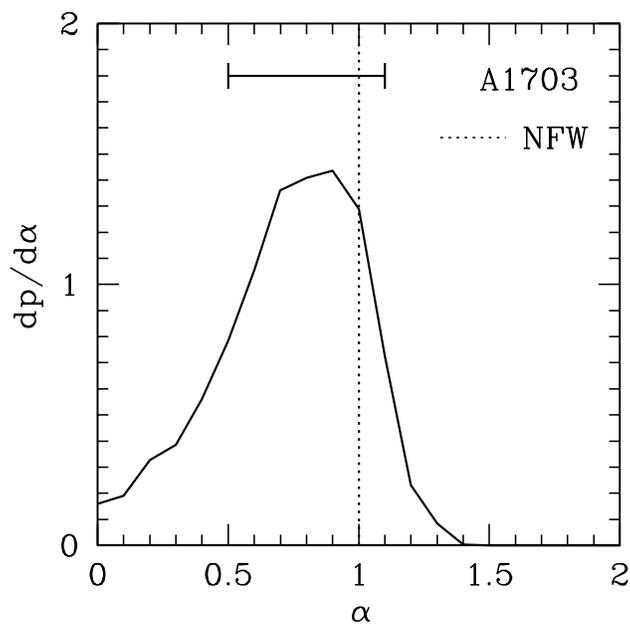}
\caption{Normalized probability distribution of the inner slope of
the dark matter density profile, $\alpha$ (see eq. [\ref{eq:gnfw}]),
from strong lens modeling of A1703. Other model parameters are
marginalized over. The vertical dotted line indicates the slope of the
original NFW profile ($\alpha=1$) which we assumed for analysis in
\S\ref{sec:slens}. The horizontal bar denotes the $1\sigma$
statistical error on $\alpha$ derived from the probability distribution.  
  \label{fig:gnfw_alpha}} 
\end{figure}

The multiple strong lens systems in A1703 for a wide range of source
redshifts constrain enclosing masses at different radii. Thus we
expect the dark matter density profile of this cluster can well be
constrained \citep{limousin08}. By using the technique described in
\S\ref{sec:slens}, we re-perform strong lens modeling of A1703 with
the dark matter distribution replaced from equation (\ref{eq:nfw}) to
a so-called generalized NFW profile \citep[e.g.,][]{jing00}:
%%%%%%%%%%%%%%%%%%
\begin{equation}
\rho(r)=\frac{\rho_s}{(r/r_s)^\alpha(1+r/r_s)^{3-\alpha}},
\label{eq:gnfw}
\end{equation}
%%%%%%%%%%%%%%%%%%
where $\alpha=1$ represent the original NFW profile. In
Figure~\ref{fig:gnfw_alpha}, we show the probability distribution of
$\alpha$ obtained from mass modeling. The result,
$\alpha=0.9_{-0.4}^{+0.2}$, is consistent with NFW, suggesting the
validity of our assumption of $\alpha=1$ in our strong lens modeling
in \S\ref{sec:slens}. An inner slope as steep as $\alpha=1.5$ is
clearly rejected.  The result is quite consistent with the
recent strong lens modeling of A1703 by \citet{richard09},
$\alpha=0.92_{-0.04}^{+0.05}$, but with much larger error. It is not
clear why our constraint on $\alpha$ is much weaker than that of 
\citet{richard09}; one of the reasons may be that we are
conservatively using only spectroscopically confirmed, robustly
identified multiple images for our mass modeling. We note that an
inner slope slightly shallower than NFW was suggested in some other
lensing clusters as well \citep{sand08}. The asymptoptic inner slope
shallower than $-1$ has also been implied by recent high-resolution
$N$-body simulations \citep[e.g.,][]{fukushige04,graham06,gao08}.

%%%%%%%%%%%%%%%%%%%%%%%%%%%%%%%%%%%%%%%%%%%%%%%%%%%%%%%%%%%%%%%%%%%%%%
\clearpage
%%%%%%%%%%%%%%%%%%%%%%%%%%%%%%%%%%%%%%%%%%%%%%%%%%%%%%%%%%%%%%%%%%%%%%
\begin{deluxetable}{ccccc}
\tablecaption{Summary of {\it Subaru} Suprime-cam images\label{tab:images}}
\tablewidth{0pt}
\tablehead{
 \colhead{Name}  & \colhead{Band} &
 \colhead{Exposure} & \colhead{Seeing}& \colhead{$m_{\rm lim}$}\\
 &  & \colhead{[sec]} & \colhead{[arcsec]}}
\startdata
A1703    & $g$ & 400$\times$3 & 0.96 & 26.39 \\
         & $r$ & 300$\times$7 & 0.80 & 26.31 \\
         & $i$ & 240$\times$5 & 0.86 & 25.58 \\
SDSS1446 & $g$ & 400$\times$3 & 0.84 & 26.22 \\
         & $r$ & 300$\times$7 & 0.82 & 26.24 \\
         & $i$ & 240$\times$5 & 0.92 & 25.47 \\
SDSS1531 & $g$ & 400$\times$3 & 0.90 & 26.28 \\
         & $r$ & 300$\times$5 & 0.98 & 26.19 \\
         & $i$ & 240$\times$5 & 1.00 & 25.38 \\
SDSS2111 & $g$ & 360$\times$4 & 0.82 & 26.13 \\
         & $r$ & 300$\times$8 & 0.60 & 26.05 \\
         & $i$ & 240$\times$7 & 0.52 & 25.48 \\
\enddata
\tablecomments{The seeing indicates the FWHM in the co-added mosaic
  image. The magnitude limit $m_{\rm lim}$ refers to the $5\sigma$
  detection limit of point sources for $2''$ aperture diameter. In
  stacking SDSS1531 $r$-band and SDSS2111 $r$-band images, we removed
  a few frames with bad seeing values; these are excluded from the
  numbers of frames indicated in this Table. 
}
\end{deluxetable}
%%%%%%%%%%%%%%%%%%%%%%%%%%%%%%%%%%%%%%%%%%%%%%%%%%%%%%%%%%%%%%%%%%%%%%%

\clearpage

%%%%%%%%%%%%%%%%%%%%%%%%%%%%%%%%%%%%%%%%%%%%%%%%%%%%%%%%%%%%%%%%%%%%%%
\begin{deluxetable}{ccrrrc}
\tablecaption{Multiple images for strong lens modeling\label{tab:strong}}
\tablewidth{0pt}
\tablehead{
 \colhead{Name}  & \colhead{Image} &
 \colhead{$\Delta x$} & \colhead{$\Delta y$} &
 \colhead{$\Delta r$} & \colhead{Redshift} \\
  & & \colhead{[arcsec]} & \colhead{[arcsec]} & 
\colhead{[arcsec]} &} 
\startdata
A1703    & A1  & $-11.7$ &  $6.8$  & 13.5 & 0.8889\\
         & A2  & $-6.3$  &  $8.4$  & 10.4 & \\
         & A3  & $-4.8$  &  $2.6$  & 5.3  & \\
         & A4  & $-10.5$ &  $0.8$  & 10.7 & \\
         & A5  & $22.1$  & $-14.9$ & 26.7 & \\
         & B1  & $-33.6$ & $-11.6$ & 35.5 & 2.627\\
         & B2  & $-25.7$ & $-21.9$ & 33.8 & \\
         & B3  & $ 21.3$ & $-28.5$ & 35.6 & \\
         & C1  & $-32.3$ & $-15.3$ & 35.7 & 2.627\\
         & C2  & $-30.4$ & $-18.0$ & 35.3 & \\
         & C3  & $21.3$  & $-28.5$ & 35.6 & \\
         & D1  & $14.1$  & $32.7$  & 35.6 & 1.908\\
         & D2  & $25.0$  & $26.7$  & 36.6 & \\
         & D3  & $-10.4$ & $37.8$  & 39.2 & \\ 
         & E1  & $-17.5$ & $32.2$  & 36.6 & 2.360\\
         & E2  & $6.9$   & $30.0$  & 30.8 & \\
         & E3  & $34.4$  & $7.2$   & 35.1 & \\
         & F1  & $20.4$  & $-18.0$ & 27.2 & 2.355\\
         & F2  & $11.0$  & $13.8$  & 17.6 & \\
         & F3  & $-36.1$ & $15.5$  & 39.3 & \\
         & F4  & $-7.1$  & $-21.3$ & 22.5 & \\
SDSS1446 & A1  & $-11.9$ & $14.9$  & 19.1 & $>1.6$\\
         & A2  & $13.2$  & $11.6$  & 17.6 &\\
         & A3  & $-4.9$  & $-10.7$ & 11.8 &\\
         & A4  & $18.9$  &  $-5.3$ & 19.6 &\\
SDSS1531 & A1  & $-11.5$ & $9.7$   & 15.0 & 1.096\\
         & A2  & $3.0$   & $-10.5$ & 10.9 &\\
         & A3  & $4.4$   &$-10.1$  & 11.0 &\\
         & A4  & $11.1$  & $-0.6$  & 11.1 &\\
         & B1  & $-13.3$ &  $5.7$  & 14.5 &1.095\\
         & B2  & $6.8$   & $3.8$   & 7.8  &\\
SDSS2111 & A1  & $10.0$  & $-5.1$  & 11.2 &\nodata\\
         & A2  & $6.1$   & $-8.0$  & 10.1 &\\
         & A3  & $-13.6$ & $-8.9$  & 16.3 &\\
         & B1  & $8.6$   & $-6.2$  & 10.6 &\nodata\\
         & B2  & $8.1$   &  $-7.1$ & 10.8 &\\
         & B3  & $-16.3$ &  $-6.8$ & 17.7 &\\
         & C1  & $12.9$  &  $-4.5$ & 13.7 &\nodata\\
         & C2  & $-0.1$  &  $-11.1$& 11.1 &\\
         & C3  & $-11.6$ &  $-9.9$ & 15.3 &\\
%%         & D1  & $-7.8$  &  $27.0$ & 28.1 &1.476\\
\enddata
\tablecomments{The same alphabet of images indicates that they are
  associated with the same source, inferred based on the colors and
  modeling. $\Delta x$ and $\Delta y$ denote the relative locations
  from the BCG of each cluster (see Figure~\ref{fig:img}); the
  positive directions correspond to West and North, respectively. 
  The distance from the BCG is shown as $\Delta r$.
  Redshifts are spectroscopic redshifts (see text for more details).
  The arc identification and spectroscopic redshifts for A1703 are
  taken from \citet{limousin08} and \citet{richard09}.} 
\end{deluxetable}
%%%%%%%%%%%%%%%%%%%%%%%%%%%%%%%%%%%%%%%%%%%%%%%%%%%%%%%%%%%%%%%%%%%%%%%

%%%%%%%%%%%%%%%%%%%%%%%%%%%%%%%%%%%%%%%%%%%%%%%%%%%%%%%%%%%%%%%%%%%%%%
\begin{deluxetable}{cccccccccc}
\rotate
\tablecaption{Result of strong lens mass modeling\label{tab:slmodel}}
\tablewidth{0pt}
\tablehead{
 \colhead{Name}  & \colhead{$M_{\rm vir}$} &
 \colhead{$x_c$} & \colhead{$y_c$} & \colhead{$e$} &
 \colhead{$\theta_e$} & \colhead{$c_{\rm vir}$} &
 \colhead{$\sigma$} & \colhead{$r_{\rm cut}$}  &
 \colhead{$\chi^2$/dof} \\ 
 & \colhead{[$10^{15}M_\odot$]} & \colhead{[arcsec]} &
 \colhead{[arcsec]} & & \colhead{[deg]} & & \colhead{[${\rm
       km\,s^{-1}}$]}  & \colhead{[arcsec]} & }  
\startdata
A1703    & $2.41_{-0.58}^{+0.70}$ & $-1.5\pm0.5$ 
         & $0.7_{-0.6}^{+0.5}$ & $0.36_{-0.03}^{+0.04}$ 
         & $-24.1_{-0.7}^{+0.6}$ & $5.1\pm0.7$ 
         & $337_{-36}^{+34}$ & $14.9\pm3.6$ 
         & 35.5/24 \\
SDSS1446 & $0.53_{-0.28}^{+1.34}$ & $1.1_{-1.6}^{+1.4}$ 
         & $-0.7_{-2.9}^{+1.9}$ & $0.42_{-0.24}^{+0.44}$ 
         & $-36.3_{-2.3}^{+2.0}$ & $11.7_{-7.2}^{+28.3}$ 
         & $231\pm38$ & $4.3_{-2.0}^{+1.9}$ 
         & 0.0/0 \\
SDSS1531 & $4.39_{-4.25}^{+2.61}$ & $0.4_{-0.5}^{+1.8}$ 
         & $-0.8_{-1.0}^{+0.3}$ & $0.10_{-0.04}^{+0.41}$ 
         & $-42.2_{-3.7}^{+7.7}$ & $3.0_{-0.7}^{+13.0}$ 
         & $291_{-24}^{+33}$ & $8.4_{-2.2}^{+2.9}$ 
         & 0.7/2 \\
%% z=1.14
%SDSS2111 & $1.03_{-0.54}^{+0.80}$ & $-3.6_{-1.9}^{+1.5}$ 
%         & $11.3_{-3.1}^{+4.8}$ & $0.36_{-0.20}^{+0.33}$ 
%         & $-168.0_{-2.2}^{+2.5}$ & $12.5_{-4.8}^{+27.5}$ 
%         & $268_{-51}^{+53}$ & $4.7_\pm2.3$ 
%         & 1.5/6 \\
%% z=1.6
SDSS2111 & $0.55_{-0.27}^{+1.30}$ & $-3.9\pm1.5$ 
         & $13.3_{-5.9}^{+3.9}$ & $0.45_{-0.26}^{+0.27}$ 
         & $-168.2_{-2.3}^{+2.7}$ & $16.0_{-10.1}^{+24.0}$ 
         & $268_{-49}^{+51}$ & $4.7_{-2.4}^{+2.6}$ 
         & 1.5/6 \\
\enddata
\tablecomments{The parameters $x_c$ and $y_c$ denote the center of a
  dark halo (NFW) component relative to the location of the BCG;
  positive $x_c$ and $y_c$ indicate the offset to West and North
  directions, respectively. The position angle $\theta_e$ is East of
  North. The values of the velocity dispersion $\sigma$ and the cutoff
  radius  $r_{\rm cut}$ in this table are those of the BCG; other member
  galaxies have values obtained from the exact scaling relation
  $\sigma\propto L^{1/4}$ and $r_{\rm cut}\propto L^{1/2}$. Note that 
  we have assumed the arc redshifts of SDSS1446 and SDSS2111 to be
  $z=1.6$ when running the MCMC; some of the best-fit values (e.g.,
  $M_{\rm vir}$)  is dependent on this assumption. }
\end{deluxetable}
%%%%%%%%%%%%%%%%%%%%%%%%%%%%%%%%%%%%%%%%%%%%%%%%%%%%%%%%%%%%%%%%%%%%%%%

%%%%%%%%%%%%%%%%%%%%%%%%%%%%%%%%%%%%%%%%%%%%%%%%%%%%%%%%%%%%%%%%%%%%%%
\begin{deluxetable}{ccc}
\tablecaption{Einstein radii from strong lens mass modeling\label{tab:slein}}
\tablewidth{0pt}
\tablehead{
 \colhead{Name}  & \colhead{$\theta_{\rm Ein}$ [arcsec]} &
 \colhead{$z_{\rm Ein}$}}
\startdata
A1703    & $12.1_{-1.3}^{+1.6}$ & 0.8889 \\
         & $22.5_{-1.7}^{+1.8}$ & 2.627  \\
SDSS1446 & $16.5_{-1.8}^{+4.7}$ & $1.6-3.5$ \\
SDSS1531 &  $9.3_{-0.8}^{+2.9}$ & 1.096 \\
%% z=1.14
%SDSS2111 & $17.7_{-4.3}^{+7.6}$ & 1.14 \\
%% z=1.6
SDSS2111 & $18.6_{-5.3}^{+4.4}$ & $<3.5$ \\
\enddata
\tablecomments{The Einstein radius $\theta_{\rm Ein}$ is that
  predicted by the best-fit dark halo component, i.e., contributions
  from member galaxies (including the BCGs) are excluded in deriving
  $\theta_{\rm Ein}$. See text for more details. } 
\end{deluxetable}
%%%%%%%%%%%%%%%%%%%%%%%%%%%%%%%%%%%%%%%%%%%%%%%%%%%%%%%%%%%%%%%%%%%%%%%

%%%%%%%%%%%%%%%%%%%%%%%%%%%%%%%%%%%%%%%%%%%%%%%%%%%%%%%%%%%%%%%%%%%%%%
\begin{deluxetable}{cccccc}
\tablecaption{Ellipticities of template stars for the PSF 
correction\label{tab:wl}}
\tablewidth{0pt}
\tablehead{
 \colhead{Name} & \colhead{$e^*_1\times10^2$}  & \colhead{$e^*_2\times10^2$} &
 \colhead{$e^*_{1,\,{\rm cor}}\times10^2$}  & \colhead{$e^*_{2,\,{\rm cor}}\times10^2$} & \colhead{$N_{\rm
 star}$} } 
\startdata
A1703    & $+0.88\pm1.11$ & $+0.56\pm1.22$ &
 $-0.02\pm0.49$ & $-0.01\pm0.43$ & 558 \\ 
SDSS1446 & $-1.08\pm0.82$ & $-0.35\pm0.76$ &
 $-0.01\pm0.45$ & $-0.02\pm0.34$ & 598 \\ 
SDSS1531 & $+0.79\pm0.94$ & $+0.14\pm 0.85$ &
 $-0.01\pm0.45$ & $+0.00\pm0.39$ & 876 \\
SDSS2111 & $-0.87\pm1.82$ & $-1.60\pm1.22$ &
 $-0.01\pm0.70$ & $+0.03\pm0.44$ & 1591 \\
\enddata
\tablecomments{Medians and standard deviations of ellipticities for
  template stars used to correct PSF anisotropies. $e^*$ and $e^*_{\rm
    cor}$ are ellipticities before and after the PSF correction. The
  last column shows the number of the template stars.} 
\end{deluxetable}
%%%%%%%%%%%%%%%%%%%%%%%%%%%%%%%%%%%%%%%%%%%%%%%%%%%%%%%%%%%%%%%%%%%%%%%

%%%%%%%%%%%%%%%%%%%%%%%%%%%%%%%%%%%%%%%%%%%%%%%%%%%%%%%%%%%%%%%%%%%%%%
\begin{deluxetable}{cccccc}
\tablecaption{Source galaxy population for weak lensing analysis\label{tab:zs}}
\tablewidth{0pt}
\tablehead{
 \colhead{Name}  & \colhead{$z_l$}  & \colhead{Magnitude}  & \colhead{$\langle
   d_{ls}/d_{os} \rangle$} & \colhead{$z_d$} & \colhead{$N_{\rm gal}$}} 
\startdata
A1703    & $0.281$& $22.0<r<25.7$ & 0.637 & 0.899 & 9155 \\
SDSS1446 & $0.464$& $22.0<r<25.3$ & 0.470 & 0.999 & 9985 \\
SDSS1531 & $0.335$& $22.0<r<24.9$ & 0.546 & 0.832 & 6032 \\
SDSS2111 & $0.637$& $22.0<r<26.0$ & 0.383 & 1.181 & 9721 \\
\enddata
\tablecomments{The range of $r$-band magnitudes, lensing depth
  $\langle  d_{ls}/d_{os} \rangle$, effective source redshifts $z_d$,
  and the numbers of source galaxies $N_{\rm gal}$ are shown. See text
  for details.}    
\end{deluxetable}
%%%%%%%%%%%%%%%%%%%%%%%%%%%%%%%%%%%%%%%%%%%%%%%%%%%%%%%%%%%%%%%%%%%%%%%

%%%%%%%%%%%%%%%%%%%%%%%%%%%%%%%%%%%%%%%%%%%%%%%%%%%%%%%%%%%%%%%%%%%%%%
\begin{deluxetable}{cccccccc}
\tablecaption{Result from weak lensing and combined strong and weak
  lensing analysis\label{tab:swmodel}}
\tablewidth{0pt}
\tablehead{ & \multicolumn{3}{c}{Weak lensing} & &
  \multicolumn{3}{c}{Strong and weak lensing}\\ 
 \cline{2-4} 
 \cline{6-8} 
 \colhead{Name}  & \colhead{$M_{\rm vir}$[$10^{15}M_\odot$]} 
 & \colhead{$c_{\rm vir}$} &  \colhead{$\chi^2$/dof} & & 
 \colhead{$M_{\rm vir}$[$10^{15}M_\odot$]} 
 & \colhead{$c_{\rm vir}$} &  \colhead{$\chi^2$/dof}
}  
\startdata
A1703    & $1.95_{-0.50}^{+0.65}$ & $3.3_{-1.1}^{+1.4}$ & 2.7/5 &
         & $1.50_{-0.35}^{+0.40}$ & $6.5_{-0.7}^{+1.2}$ & 7.9/7 \\
SDSS1446 & $0.83_{-0.25}^{+0.29}$ & $9.1_{-4.1}^{+11.4}$& 6.3/5 &
         & $0.83_{-0.22}^{+0.30}$ & $8.3_{-3.1}^{+3.9}$ & 6.4/6 \\
SDSS1531 & $0.59_{-0.26}^{+0.39}$ & $11.5_{-7.4}^{+28.5}$& 8.0/5 &
         & $0.66_{-0.24}^{+0.29}$ & $7.9_{-1.5}^{+3.0}$  & 8.1/6 \\
%% z=1.14
%SDSS2111 & $0.92_{-0.32}^{+0.41}$ & $14.1_{-9.5}^{+25.9}$& 7.5/5 &
%         & $0.92_{-0.30}^{+0.37}$ & $14.1_{-6.2}^{+24.8}$& 7.5/6 \\
%% z=1.6
SDSS2111 & $0.92_{-0.32}^{+0.41}$ & $14.1_{-9.5}^{+25.9}$& 7.5/5 &
         & $0.92_{-0.32}^{+0.41}$ & $14.1_{-9.3}^{+25.9}$& 7.5/6 \\
\enddata
\tablecomments{Weak lensing constraints come from radial profiles of
  tangential shear $g_+$ (see Figure~\ref{fig:profile_gamma}). The
  Einstein radii $\theta_{\rm Ein}$ inferred from strong lens modeling
  are used as strong lens constraints (see Table~\ref{tab:slein}).
  Note that we have restricted the range of the concentration parameter
  to $c_{\rm vir}<40$. The virial radii of the best-fit models from
  strong and weak lensing analysis are $2.56$, $1.91$, $1.89$, and
  $1.80$~Mpc for A1703, SDSS1446, SDSS1531, and SDSS2111, respectively.}
\end{deluxetable}
%%%%%%%%%%%%%%%%%%%%%%%%%%%%%%%%%%%%%%%%%%%%%%%%%%%%%%%%%%%%%%%%%%%%%%%

\clearpage

%%%%%%%%%%%%%%%%%%%%%%%%%%%%%%%%%%%%%%%%%%%%%%%%%%%%%%%%%%%%%%%%%%%%%%%
%%%%%%%%%%%%%%%%%%%%%%%%%%%%%%%%%%%%%%%%%%%%%%%%%%%%%%%%%%%%%%%%%%%%%%%

%%%%%%%%%%%%%%%%%%%%%%%%%%%%%%%%%%%%%%%%%%%%%%%%%%%%%%%%%%%%%%%%%%%%%%%
%%%%%%%%%%%%%%%%%%%%%%%%%%%%%%%%%%%%%%%%%%%%%%%%%%%%%%%%%%%%%%%%%%%%%%%

%%%%%%%%%%%%%%%%%%%%%%%%%%%%%%%%%%%%%%%%%%%%%%%%%%%%%%%%%%%%%%%%%%%%%%%
%%%%%%%%%%%%%%%%%%%%%%%%%%%%%%%%%%%%%%%%%%%%%%%%%%%%%%%%%%%%%%%%%%%%%%%

%%%%%%%%%%%%%%%%%%%%%%%%%%%%%%%%%%%%%%%%%%%%%%%%%%%%%%%%%%%%%%%%%%%%%%%
%%%%%%%%%%%%%%%%%%%%%%%%%%%%%%%%%%%%%%%%%%%%%%%%%%%%%%%%%%%%%%%%%%%%%%%

%%%%%%%%%%%%%%%%%%%%%%%%%%%%%%%%%%%%%%%%%%%%%%%%%%%%%%%%%%%%%%%%%%%%%%%
%%%%%%%%%%%%%%%%%%%%%%%%%%%%%%%%%%%%%%%%%%%%%%%%%%%%%%%%%%%%%%%%%%%%%%%

%%%%%%%%%%%%%%%%%%%%%%%%%%%%%%%%%%%%%%%%%%%%%%%%%%%%%%%%%%%%%%%%%%%%%%%
%%%%%%%%%%%%%%%%%%%%%%%%%%%%%%%%%%%%%%%%%%%%%%%%%%%%%%%%%%%%%%%%%%%%%%%

%%%%%%%%%%%%%%%%%%%%%%%%%%%%%%%%%%%%%%%%%%%%%%%%%%%%%%%%%%%%%%%%%%%%%%%
%%%%%%%%%%%%%%%%%%%%%%%%%%%%%%%%%%%%%%%%%%%%%%%%%%%%%%%%%%%%%%%%%%%%%%%

%%%%%%%%%%%%%%%%%%%%%%%%%%%%%%%%%%%%%%%%%%%%%%%%%%%%%%%%%%%%%%%%%%%%%%%
%%%%%%%%%%%%%%%%%%%%%%%%%%%%%%%%%%%%%%%%%%%%%%%%%%%%%%%%%%%%%%%%%%%%%%%

%%%%%%%%%%%%%%%%%%%%%%%%%%%%%%%%%%%%%%%%%%%%%%%%%%%%%%%%%%%%%%%%%%%%%%%
%%%%%%%%%%%%%%%%%%%%%%%%%%%%%%%%%%%%%%%%%%%%%%%%%%%%%%%%%%%%%%%%%%%%%%%

\end{document}